\documentclass[journal]{IEEEtran}
\usepackage{amsbsy}
\usepackage{floatflt} 

\usepackage{amsmath}
\usepackage{amssymb}
\usepackage{times}
\usepackage{graphics}
\usepackage{graphicx}
\usepackage{xspace}
\usepackage{paralist} 
\usepackage{setspace} 
\usepackage{xypic}
\xyoption{curve}
\usepackage{latexsym}
\usepackage{theorem}
\usepackage{ifthen}
\usepackage{caption}
\usepackage{subfigure}
\usepackage{booktabs}
\usepackage{algorithm}
\usepackage{algorithmic}
\usepackage{color}
\newcommand{\ra}[1]{\renewcommand{\arraystretch}{#1}}
\ifCLASSINFOpdf
\else
\fi

{\theoremheaderfont{\it} \theorembodyfont{\rmfamily}
\newtheorem{theorem}{Theorem}
\newtheorem{lemma}[theorem]{Lemma}

}

\newcounter{brojac}

{\theoremheaderfont{\it} \theorembodyfont{\rmfamily}
\newtheorem{assumption}[brojac]{Assumption}
}

\newcounter{brojac2}

{\theoremheaderfont{\it} \theorembodyfont{\rmfamily}
\newtheorem{remark}[brojac2]{Remark}
}

\begin{document}
\title{Distributed Gradient Methods with Variable Number of Working Nodes}

\author{Du$\check{\mbox{s}}$an Jakoveti\'c, Dragana Bajovi\'c, Nata$\check{\mbox{s}}$a Kreji\'c, and Nata$\check{\mbox{s}}$a Krklec Jerinki\'c
\thanks{D. Jakoveti\'c and D. Bajovi\'c are with University of Novi Sad, BioSense Institute, Novi Sad, Serbia.
 N. Kreji\'c and N. Krklec Jerinki\'c are with Department of Mathematics and Informatics, Faculty of Science, University of Novi Sad, Novi Sad, Serbia. Research of N. Kreji\'c and N. Krklec-Jerinki\'c is supported by Ministry of Education, Science and Technological Development, Republic of Serbia, grant no. 174030. Authors' e-mails: [djakovet,dbajovic]@uns.ac.rs, natasak@uns.ac.rs,
  natasa.krklec@dmi.uns.ac.rs.}}
%
%


\maketitle

\begin{abstract}
We consider distributed optimization where $N$ nodes in a connected network minimize
the sum of their local costs subject to a common constraint set.
We propose a distributed projected gradient method where each node, at each iteration~$k$,
performs an update (is active) with probability~$p_k$,
and stays idle (is inactive) with probability~$1-p_k$. Whenever active, each node performs an update by
 weight-averaging its solution estimate with the estimates of its active neighbors,
taking a negative gradient step with respect to its local cost, and performing a projection
onto the constraint set; inactive nodes perform no updates.
Assuming that nodes' local costs are strongly convex, with Lipschitz continuous gradients, we show that, as long as
activation probability $p_k$ grows to one asymptotically,
our algorithm converges in the mean square sense (MSS) to the same solution as the standard distributed gradient method,
i.e., as if all the nodes were active at all iterations. Moreover, when $p_k$ grows to one linearly,
with an appropriately set convergence factor,
the algorithm has a linear MSS convergence, with practically the same factor as the standard distributed gradient method.
Simulations on both synthetic and real world data sets
demonstrate that, when compared with the standard distributed gradient method,
 the proposed algorithm significantly reduces
the overall number of per-node communications and per-node gradient evaluations (computational cost)
 for the same required accuracy.
\end{abstract}


\maketitle \thispagestyle{empty} \maketitle
%
%
%
%
\vspace{-1mm}
\section{Introduction}
\label{section-introduction}
%
%
%
%
%
%
%
We consider distributed optimization where $N$ nodes
constitute a generic, connected network, each node $i$
 has a convex cost function $f_i\,: {\mathbb R}^d \mapsto {\mathbb R}$ known
 only by $i$, and the nodes want to solve the following problem:
\begin{equation}
\label{eqn-opt-prob-original}
\begin{array}[+]{ll}
\mbox{minimize} \:\:\sum_{i=1}^N f_i(x) =:f(x)\\
\mbox{subject to}\:\: x \in \mathcal{X}.
\end{array}
\end{equation}
Here, $x \in {\mathbb R}^d$ is the optimization variable common to all nodes,
and $\mathcal X \subset {\mathbb R}^d$
\nocite{ScutariBigData,GiannakisBigData1,GiannakisBigData2,RibeiroADMM1,Rabbat,SoummyaEst,SayedConf,SayedEstimation,NecoaraMPC,JoaoMotaMPC}
 is a closed, convex constraint set, known by all.
The above and related problems arise frequently, e.g., in
big data analytics in cluster or cloud environments, e.g.,~\cite{ScutariBigData}-\cite{GiannakisBigData2},
distributed estimation in wireless sensor networks (WSNs), e.g.,~\cite{RibeiroADMM1}-\cite{SayedEstimation},
and distributed control applications, e.g.,~\cite{NecoaraMPC,JoaoMotaMPC}.
With all the above applications, data is split across
multiple networked nodes (sensors, cluster machines, etc.), and
$f_i(x) = f_i(x; D_i)$ represents
a loss with respect to data~$D_i$ stored locally at node~$i$.

A popular approach to solve~\eqref{eqn-opt-prob-original} is via distributed (projected) (sub)gradient methods,
e.g.,~\cite{nedic_T-AC,nedic_novo,nedic-gossip}. With these methods, each node~$i$,
 at each iteration~$k$, updates its solution estimate by weight-averaging it with
 the estimates of its neighbors, taking a negative gradient
 step with respect to its local cost, and projecting the result onto the constraint set~$\mathcal X$.
 Distributed gradient methods are attractive as they do not require centralized coordination,
 have inexpensive iterations (provided that projections onto $\mathcal X$ are computationally light),
 and exhibit resilience to inter-node communication failures and delays; however, they
 have a drawback of slow convergence rate. 


Several techniques to improve convergence rates
of distributed (projected) gradient methods have been
proposed, including Newton-like methods, e.g.,~\cite{NetworkNewtonPart1,NetworkNewtonPart2,NetworkNewton,NewtonAsu},
and Nesterov-like methods, e.g.,~\cite{arxivVersion,AnnieChen}.
In this paper, we make distributed (projected) gradient methods
more efficient by proposing a novel method with
a variable number of working nodes. Each node~$i$, at each iteration~$k$,
performs an update (is active) with probability~$p_k$, and stays idle (is inactive)
 with probability~$1-p_k$, while the activation
 schedule is independent both across nodes and across iterations. Whenever active, each node~$i$ performs the same update as
 with the standard distributed gradient method, while inactive nodes perform no updates.

Our main results are as follows. Assuming that the costs $f_i$'s are strongly convex and their gradients are Lipschitz
 continuous, we show that, whenever the activation probability~$p_k$
  grows asymptotically to one, our method converges in the mean square sense
  to the same point as the standard distributed gradient method.\footnote{Under a constant
  step-size $\alpha$, the standard (projected) distributed gradient method
   converges to a point in a neighborhood
   of the solution of~\eqref{eqn-opt-prob-original},
   where the corresponding squared distance is $O(\alpha)$;
   see ahead Theorem~\ref{theorem-standard-dis-grad-generic} and, e.g., \cite{WotaoYinDisGrad,Matei}.}
   If, in addition, $p_k$ converges to unity at
   a sublinear rate at least as fast as $1/k^{1+\zeta}$
   ($\zeta>0$ arbitrarily small), the method also
   converges almost surely.

Our second group of results assumes that $p_k$ grows to one linearly, with the convergence factor~$\delta \in (0,1)$.
   In this setting, we show that the proposed algorithm has a linear convergence rate (in the sense of the expected distance to the solution). When, in addition, quantity $\delta $ is set in accordance with the
    $f_i$'s condition number, we show that
    the proposed algorithm converges practically with the same linear convergence
    factor as the standard distributed gradient method (albeit with a larger hidden constant). Hence, interestingly,
    our algorithm achieves practically the same rate in iterations~$k$
    as the standard distributed gradient method, but with the reduced cost per iteration~$k$ (overall
    communication and computational cost), thus making distributed gradient methods more efficient.
     Simulation examples on $l_2$-regularized logistic losses -- both
     on synthetic and real world data sets --
     confirm that our method significantly reduces the communication and computational
     costs with respect to the standard distributed gradient method, for the same desired accuracy.
        Simulations also demonstrate that the proposed method
      exhibits a high degree of resilience to asynchrony.

      Communication and computational savings are highly relevant with applications like
     WSNs and distributed learning in cluster or cloud environments.
      With WSNs, the reduced communication and computational cost
      to retrieve the result translate into energy saving of the
      sensor motes' batteries and the increase of the network lifetime.
      With distributed learning in cluster or cloud environments, less amount
      of communication and computation for a
      specific application/task means that the saved resources
      can be re-allocated to another concurrent tasks. For example,
       at times when a node with our method is idle,
      the resources allocated to it (e.g., a virtual
      cloud machine) can be released and re-allocated to other tasks.

We explain intuitively the above results that we achieve. Namely, standard distributed gradient method
exhibits, in a sense, two sources of redundancy--the first corresponds to the
inter-node communications aspect, while the second corresponds to an optimization aspect
(number of gradient evaluations per iteration)
 of the algorithm.
 It turns out that, as we show here, a careful simultaneous exploitation of these two redundancies
allows to match the rate of the standard distributed gradient method with reduced communications
and computational costs.
The two sources of redundancy have been already noted in the literature but
have not been exploited simultaneously before. The communication redundancy, e.g.,~\cite{asu-random}
 means that the inter-node communications can be ``sparsified,'' e.g., through
 a randomized protocol, so that the algorithm still remains convergent. In other words, it is
not necessary to utilize communications through all the available links at all iterations
for the algorithm to converge. The optimization redundancy
 has been previously studied only in the context of
 centralized optimization, e.g.,~\cite{SchmidtHybrid}.
   Increasing
the probability~$p_k$ here resembles increasing the size of the batch
 used along iterations in~\cite{SchmidtHybrid}.
 The core idea is that, under certain assumptions on the cost functions, a (centralized) stochastic-type gradient method with
an appropriately increasing sample size matches the convergence rate of the
standard gradient method with the full sample size at all iterations, as shown in~\cite{SchmidtHybrid}.
 More broadly, the idea to use ``increasingly-accurate'' optimization
 steps in iterative optimization algorithms has been
 widely considered in the (centralized) optimization literature,
 including, e.g., inexact Newton methods~\cite{InexactNewton1}.

Finally, although the focus of the paper is on
the standard distributed gradient method~\cite{nedic_T-AC-private},
we believe that the proposed idling strategy is of a more general applicability. An interesting future research direction is
to incorporate it in other distributed methods, including, e.g.,
     \cite{SayedOptim,arxivVersion,AnnieChen,EXTRA,NetworkNewtonPart1}.

\subsection{Related work}
\label{subsection-related-work}
We now briefly review existing work relevant to our contributions
 to help us further contrast our work from the literature.
We divide the literature into two classes: 1) distributed gradient methods for multi-agent optimization;
 and 2) centralized stochastic approximation methods with variable sample sizes.
 The former class relates to our work through the communication redundancy, while the latter
 considers the optimization redundancy.

\textbf{Distributed gradient methods for multi-agent optimization}.
 Distributed methods of this type date back at least to the 80s, e.g.,~\cite{Tsitsiklis_Distr_Opt},
 and have received renewed interest in the past decade, e.g.,~\cite{nedic_T-AC}.
 Reference~\cite{nedic_T-AC} proposes the distributed (sub)gradient
 method with a constant step-size, and analyzes its
 performance under time-varying communication networks.
 Reference~\cite{asu-random} considers distributed (sub)gradient methods
 under random communication networks with failing links and establishes
 almost sure convergence under a diminishing step-size rule.
 A major difference of our paper from the above works is
 that, in \cite{Tsitsiklis_Distr_Opt,nedic_T-AC,asu-random},
 only inter-node communications over iterations are ``sparsified,''
  while each node performs gradient evaluations at each iteration~$k$.
 In~\cite{nedic-gossip}, the authors propose a gossip-like scheme where, at each~$k$,
 only two neighboring nodes in the network wake up and perform
 weight-averaging (communication among them) and a negative gradient step
 with respect to their respective local costs, while the remaining nodes stay idle.
  The key difference with respect to our paper is that, with our method,
  the number of active nodes over iterations~$k$ (on average) is increasing,
  while in~\cite{nedic-gossip} it remains equal to two for all~$k$. Consequently,
  the established convergence properties of the two methods are very different.

There have been many works where nodes or links in
the network are controlled by random variables.
References~\cite{SoummyaTopologyDesign,DimakisBoradcast,Scaglione,SouumyaQuantized}
 consider distributed algorithms
 for solving the consensus problem -- finding
 an average of nodes' local scalars $a_i$'s,
 while we consider here a more general problem~\eqref{eqn-opt-prob-original}.
 These consensus algorithms involve
  only local averaging steps, and
  no local gradient steps are present (while we have
  here both local averaging and local gradient steps).
  The models of averaging (weight) matrices
  which~\cite{SoummyaTopologyDesign,DimakisBoradcast,Scaglione,SouumyaQuantized}
  assume are very different from ours: they all
  assume random weight matrices with time-invariant
  distributions, while ours are time-varying.
  Reference~\cite{SayedChangingTopIcassp}
   studies diffusion algorithms under
   changing topologies and data-normalized algorithms, under general, non-Gaussian distributions.
Reference~\cite{TakahashiYamadaLinkProbContr}
proposes a control mechanism for link activations in diffusion algorithms
  to minimize the estimation error under given resource constraints.
  The main differences with respect to our paper
  are that~\cite{TakahashiYamadaLinkProbContr}
   assumes that local gradients are always
   incorporated (deterministic step-sizes),
   and the link activation probabilities are time invariant.

References~\cite{SayedAsynch1,SayedAsynch2,SayedAsynch3}
 provide a thorough and in-depth analysis
 of diffusion algorithms under a very general
  model of asynchrony, where
  both the combination (weight)
   matrices and nodes' step-sizes are random.
   Our work differs from these references in several aspects,
   which include the following. A major difference is that
   papers~\cite{SayedAsynch1,SayedAsynch2,SayedAsynch3}
    assume that both the step sizes'and the combination matrices'
     random processes have constant (time-invariant)
      first and second moments, and the two processes
      are moreover mutually independent.
      In contrast, both our
      weight matrices and step-sizes have time-varying distributions.
      Actually, the fact that, with our method, node activation probabilities
      converge to one (which corresponds to the
      time-varying first moment of the step-sizes) is critical
      to establish our main results~(Theorems~2 and~3).
 Further differences are that papers~\cite{SayedAsynch1,SayedAsynch2,SayedAsynch3} allow
      for noisy gradients, their nodes' local cost functions
      all have the same minimizers, and therein the optimization problem is unconstrained.
      In contrast, we assume noise-free gradients, different local minimizers,
      and constrained problems.

Our paper is also related to reference~\cite{SayedInformedAgents}, which
 considers diffusion algorithms
 with two types of nodes -- informed and uninformed. The informed
 nodes both: 1) acquire measurements and perform in-network processing
  (which translates into computing gradients in our scenario); and 2)
 perform consultation with neighbors (which translates
  into weight-averaging the estimates across neighborhoods),
  while the uninformed nodes only perform the latter task.
  The authors study the effect of the proportion of informed nodes and their distribution in space.
  A key difference with respect to our work is that
  the uninformed nodes in~\cite{SayedInformedAgents} still perform
  weight-averaging, while the idle nodes here perform no processing.
   Finally, we comment on reference~\cite{RabbatAdaptive}
  which introduces an adaptive
  policy for each node to decide whether
  it will communicate with its neighbors or not and
  demonstrate significant savings in communications
  with respect to the always-communicating scenario.
   A major difference of~\cite{RabbatAdaptive} from our paper is
  that, with~\cite{RabbatAdaptive}, nodes always perform local gradients, i.e., they
  do not stay idle (in the sense defined here).


\textbf{Centralized stochastic approximation methods with variable sample sizes} have
been studied for a long time. We distinguish two types of methods: the ones that
assume unbounded sample sizes (where the cost function
is in the form of a mathematical expectation) and the methods with bounded sample sizes
 (where the cost function is of the form in~\eqref{eqn-opt-prob-original}.)
 Our work contrasts with both of these threads of works by considering
 distributed optimization over an arbitrary connected network, while they consider
 centralized methods.

Unbounded sample sizes have been studied, e.g., in~\cite{Feris,Tito,Polak,P1,P2}. Reference~\cite{Feris} uses
 a Bayesian scheme to determine the sample size at each iteration within the trust region framework,
 and it shows almost sure convergence to a problem solution.
 Reference~\cite{Tito} shows almost sure convergence as long as the sample size grows
 sufficiently fast along iterations. In~\cite{Polak}, the variable sample size strategy
 is obtained as the solution of an associated auxiliary optimization problem.
 Further references on careful analyses of the increasing sample sizes are, e.g.,~\cite{P1,P2}.


References~\cite{Bastin,BTuan} consider a trust region framework
and assume bounded sample sizes,
but, differently from our paper and~\cite{SchmidtHybrid,Feris,Polak,P1,P2}, they allow the sample size
both to increase and to decrease at each iteration. The paper
chooses a sample size at each iteration such that a balance is achieved between
the decrease of the cost function and the width of an associated confidence interval.
 Reference~\cite{nas} proposes a schedule sequence in the monotone line search framework which
 also allows the sample size both increase and decrease at each iteration;
  paper~\cite{nas2} extends the results in~\cite{nas} to a non-monotone line search.

Reference~\cite{SchmidtHybrid} is closest to our paper
within this thread of works, and
our work mainly draws inspiration from it. The authors consider a bounded sample size, as we do here.
 They consider both deterministic and
stochastic sampling and determine the increase
of the sample size along iterations such that the algorithm
attains (almost) the same rate as if the full sample size was used at all iterations.
 A major difference of~\cite{SchmidtHybrid} with respect to the current paper
 is that they are not concerned with the networked scenario,
 i.e., therein a central entity works with the variable (increasing) sample size.
 This setup is very different from ours as it has no problem dimension of propagating information
 across the networked nodes -- the dimension present in distributed multi-agent optimization.


\textbf{Paper organization}.
 The next paragraph introduces notation. Section~{II}
  explains the model that we assume and presents our proposed distributed algorithm.
  Section~{III} states our main results which we prove in Section~{IV}.
  Section~{V} provides numerical examples.
  Section~{VI} provides a discussion on the results
  and gives extensions. Finally, we conclude in Section~{VII}.
    Certain auxiliary proofs are provided in the Appendix.

\textbf{Notation}. We denote by: $\mathbb R$ the set of real numbers; ${\mathbb R}^d$ the $d$-dimensional
Euclidean real coordinate space; $A_{ij}$ the entry in the $i$-th row and $j$-th column of a matrix $A$;
$A^\top$ the transpose of a matrix $A$; $\odot$
and $\otimes$ the Hadamard (entry-wise) and Kronecker product of matrices, respectively; $I$, $0$, $\mathbf{1}$, and $e_i$, respectively, the identity matrix, the zero matrix, the column vector with unit entries, and the $i$-th column of $I$; $J$ the $N \times N$ matrix $J:=(1/N)\mathbf{1}\mathbf{1}^\top$;
$A \succ  0 \,(A \succeq  0 )$ means that
the symmetric matrix $A$ is positive definite (respectively, positive semi-definite);
 $\|\cdot\|=\|\cdot\|_2$ the Euclidean (respectively, spectral) norm of its vector (respectively, matrix) argument; $\lambda_i(\cdot)$ the $i$-th largest eigenvalue, $\mathrm{Diag}\left(a\right)$ the diagonal matrix with the diagonal equal to the vector $a$; $|\cdot|$ the cardinality of a set; $\nabla h(w)$ the gradient evaluated at $w$ of a function $h: {\mathbb R}^d \rightarrow {\mathbb R}$, $d \geq 1$; $\mathbb P(\mathcal A)$ and $\mathbb E[u]$ the probability of
an event $\mathcal A$ and expectation of a random variable $u$, respectively.
For two positive sequences $\eta_n$ and $\chi_n$, we have: $\eta_n = O(\chi_n)$ if
 $\limsup_{n \rightarrow \infty}\frac{\eta_n}{\chi_n}<\infty$.
 Finally, for a positive sequence $\chi_n$
  and arbitrary sequence $\xi_n$,
  $\xi_n = o(\chi_n)$ if $\lim_{n \rightarrow \infty}\frac{\xi_n}{\chi_n}=0$.

 %
 %
 %
 %
 %
%
%
%
%
\section{Model and algorithm}
\label{section-model-and-preliminaries}
Subsection \ref{subsection-opt-model}
describes the optimization and network models
that we assume, while Subsection~\ref{subsection-proposed-algorithm}
presents our proposed distributed algorithm with
variable number of working nodes.

\subsection{Problem model}
\label{subsection-opt-model}

\textbf{Optimization model}. We consider optimization problem~\eqref{eqn-opt-prob-original},
%
%
 and we impose the following assumptions on~\eqref{eqn-opt-prob-original}.

\begin{assumption}[Optimization model]
\label{assumption-f-i-s}
\begin{enumerate}[(a)]
\item For all $i$, $f_i\,: {\mathbb R}^d \mapsto {\mathbb R}$ is strongly convex with modulus $\mu>0$, i.e.:
\[
f_i(y) \geq f_i(x) + \nabla f_i(x)^\top (y-x) + \frac{\mu}{2}\|y-x\|^2, \:\: \forall x,y \in {\mathbb R}^d.
\]
\item
For all $i$, $f_i\,: {\mathbb R}^d \mapsto {\mathbb R}$
has Lipschitz continuous gradient with constant $L$, $0< \mu \leq L < \infty$, i.e.:
\[
\|\nabla f_i(x) - \nabla f_i(y)\| \leq L \|x-y\|,\:\:\:\forall x,y \in {\mathbb R}^d.
\]
\item The set $\mathcal X \subset {\mathbb R}^d$ is nonempty, closed, convex, and bounded.
\end{enumerate}
\end{assumption}
We denote by $D:=\max\{\|x\|:\, x \in \mathcal X\}$ the diameter of $\mathcal X$.
 Note that, as $\mathcal X$ is compact,
 the gradients $\nabla f_i(x)$'s are bounded over $x \in \mathcal X$, i.e.,
 there exists $G>0$, such that,
 for all $i,$ for all $x \in \mathcal X$,
 $\|\nabla f_i(x)\| \leq G$.
 The constant $G$
  can be taken as $L D + \max_{i=1,...,N}\|\nabla f_i(0)\|$. Indeed, for
  any $x \in \mathcal X$, we have:
 $
 \|\nabla f_i(x)\| \leq  \|\nabla f_i(x) - \nabla f_i(0)\| + \|\nabla f_i(0)\| $
 $\leq$ $
 L \|x\| + \|\nabla f_i(0)\| $ $
 \leq$ $ L\,D +  \max_{i=1,...,N}\|\nabla f_i(0)\|.
 $
 Similarly, there exist constants $-\infty < m_f \leq M_f < \infty$, such that
 $m_f \leq f_i(x) \leq M_f$, $\forall i$, $\forall x \in \mathcal X$.
 Constants $m_f$ and $M_f$ can be taken
 as $M_f = -m_f = G\,D + \max_{i=1,...,N}|f_i(0)|$.
 Under Assumption~\ref{assumption-f-i-s},
 \eqref{eqn-opt-prob-original} is solvable and has a unique solution, which
 we denote by $x^\star$.

\textbf{Network model}. Nodes are connected in a generic undirected network
$\mathcal G = (\mathcal{V}, E)$, where $\mathcal V$ is the set of $N$
 nodes and $E$ is the set of edges -- all node (unordered) pairs $\{i,j\}$
 that can exchange messages through a communication link.
 We impose the following assumption.

\begin{assumption}[Network connectedness]
\label{assumption-network-connectedness}
The network $\mathcal G = (\mathcal{V}, E)$ is connected, undirected, and simple (no self-loops nor multiple links).
\end{assumption}
Both Assumptions~1 and~2 hold throughout the paper.
We denote by $\Omega_i$ the neighborhood set of node $i$~(excluding $i$).
 We associate with $\mathcal G$ a
 $N \times N$ symmetric weight matrix $C$,
 which is also stochastic (rows sum to one and all the entries are non-negative).
 We let $C_{ij}$ be strictly
 positive for each $\{i,j\} \in E$, $i \neq j$;
  $C_{ij}=0$ for $\{i,j\} \notin E$, $i \neq j$;
  and $C_{ii} = 1-\sum_{j \neq i}C_{ij}$.
 As we will see, the weights $C_{ij}$'s will play a role in
 our distributed algorithm. The quantities $C_{ij}$, $j \in \Omega_i$,
 are assumed available to node~$i$ before
 execution of the distributed algorithm.
  We assume that matrix~$C$ has strictly positive
  diagonal entries (each node assigns a
  non-zero weight to itself) and is positive definite, i.e., $\lambda_N(C)>0$.
 For a given arbitrary stochastic, symmetric weight matrix $C^\prime$ with
  positive diagonal elements, positive definiteness may not hold.
  However, such arbitrary~$C^\prime$ can be easily adapted
  to generate matrix~$C$ that obeys all the
  required properties (symmetric, stochastic, positive diagonal elements),
  and, in addition, is positive definite. Namely, letting, for some~$\kappa \in (0,1)$,
  $C: = \frac{\kappa+1}{2} I + \frac{1-\kappa}{2} C^\prime$,
  we obtain that $\lambda_N(C) > \kappa$.
   It can be shown that, under the above assumptions on $C$,
  $\lambda_1(C)=1$, and  $\lambda_2(C) <1.$

Note that, in the assumed model, matrix~$C$ is \emph{doubly
stochastic}, i.e., not only its rows but also its
columns sum up to one. It is worth noting that, for practical implementations,
it is certainly relevant to consider matrices $C$
 which are only \emph{row-stochastic}, and not also
 \emph{column-stochastic}. It is an
 interesting future research direction to
 extend the proposed method to row-stochastic matrices also, relying
 on prior works including, e.g.,~\cite{NedicPush}.


\subsection{Proposed distributed algorithm}
\label{subsection-proposed-algorithm}
We now describe
the distributed algorithm to solve~\eqref{eqn-opt-prob-original}
 that we propose. We assume that all nodes are synchronized according
 to a global clock and simultaneously (in parallel) perform iterations
 $k=0,1,...$ At each iteration~$k$, each node $i$ updates its solution
 estimate~$x_i^{(k)} \in {\mathcal X}$, with arbitrary
 initialization~$x_i^{(0)} \in \mathcal X$.
  To avoid notational clutter, we
 will assume that $x_i^{(0)}=x_j^{(0)}$, $\forall i,j$.
   Further,
 each node has an internal Bernoulli state variable $z_i^{(k)}$.
 If $z_i^{(k)}=1$, node $i$ updates $x_i(k)$ at iteration $k$;
 we say that, in this case, node $i$ is active at~$k$.
 If $z_i^{(k)}=0$, node $i$ keeps its current state $x_i(k)$ and does not
 perform an update; we say that, in this case, node $i$ is idle.
 At each $k$, each node $i$ generates $z_i^{(k)}$
  independently from the previous iterations, and independently from other nodes.
   We denote by $p_k: = \mathbb P \left( z_i(k) = 1\right)$.
   The quantity $p_k$ is our algorithm's tuning parameter,
   and is common for all nodes.
   We assume that, for all $k$, $p_k \geq p_{\mathrm{min}},$
    for a positive constant $p_{\mathrm{min}}.$

Denote by~$\Omega_i^{(k)}$ the set of working
neighbors of node $i$ at $k$, i.e.,
all nodes $j \in \Omega_i$ with $z_j^{(k)}=1$.
 The update of node $i$ is as follows. If $z_i^{(k)}=0$,
 node $i$ is idle and sets $x_i^{(k+1)} = x_i^{(k)}$. Otherwise, if $z_i^{(k)}=1$,
node $i$ broadcasts its state to all its working neighbors
$j \in \Omega_i^{(k)}$. The
non-working (idle) neighbors do not receive $x_i^{(k)}$; for
example, with WSNs, this corresponds to
switching-off the receiving antenna of a node.
   Likewise, node $i$
  receives~$x_j^{(k)}$ from all $j \in \Omega_i^{(k)}$.
  Upon reception, node $i$ updates $x_i^{(k)}$ as follows:
  \begin{eqnarray}
  \label{eqn-update-one-node}
  x_i^{(k+1)} &=& \mathcal{P}_{\mathcal X} \left\{ \,\left( 1-\sum_{j \in \Omega_i^{(k)}} C_{ij} \right) x_i^{(k)} \right.\\
  &+& \left. \sum_{j \in \Omega_i^{(k)}} C_{ij} \, x_j^{(k)} - \frac{\alpha}{p_k} \nabla f_i(x_i^{(k)})\,\right\}. \nonumber
  \end{eqnarray}
  In~\eqref{eqn-update-one-node},
  $\mathcal{P}_{\mathcal X}(y)=\mathrm{arg\,min}_{v \in \mathcal X}\|v-y\|$
   denotes the Euclidean projection of point~$y$ on $\mathcal X$,
   and $\alpha>0$ is a constant;
   we let $\alpha \leq \lambda_N(C)/L$. (See ahead Remark~\ref{remark-1}.) In words,
  \eqref{eqn-update-one-node} means that node $i$
   makes a convex combination of its own estimate
   with the estimates of its working neighbors, takes a step in the negative direction of its local gradient,
   and projects the resulting value onto the constraint set.
    As we will see, multiplying the step-size in~\eqref{eqn-update-one-node} by $1/p_k$
     compensates for non-working (idle) nodes over iterations.

\begin{remark}
Setting $p_k = 1$, $\forall k$, corresponds to
the standard distributed (sub)gradient method in~\cite{nedic_T-AC-private}.
\end{remark}


\textbf{Compact representation}.
We present~\eqref{eqn-update-one-node} in a compact form.
Denote by $x^{(k)}:=(\,(x_1^{(k)})^\top,...,(x_N^{(k)})^\top)^\top$, 
and $z^{(k)}:=(z_1^{(k)},...,z_N^{(k)})^\top$.
Further, introduce $F: {\mathbb R}^{N\,d} \mapsto \mathbb R$,
with
\[
F(x)=F(x_1,...,x_N):=\sum_{i=1}^N f_i(x_i).
\]
 Also, denote by $\mathcal X^N \subset {\mathbb R}^{N\,d}$ the Cartesian product
 $\mathcal X \times ... \times \mathcal X$, where $\mathcal X$ is repeated $N$ times.
Next, introduce the $N \times N$ random matrix $W^{(k)}$,
defined as follows:
\[
W^{(k)}_{ij} =
\left\{ \begin{array}{lll}
 C_{ij} z_i^{(k)} z_j^{(k)} &\mbox{ for $\{i,j\} \in E$, $i \neq j$} \\
 0 &\mbox{ for $\{i,j\} \notin E$, $i \neq j$} \\
 1-\sum_{s \neq i} W_{is}^{(k)} &\mbox{ for $i=j$.}
       \end{array} \right.
       \]
       %
Then, it is easy to see that, for $k=0,1,...,$ update rule~\eqref{eqn-update-one-node}
 can be written as:
 \begin{eqnarray}
 \label{eqn-update-compact}
 x^{(k+1)} &=& \mathcal{P}_{\mathcal X^N}\left\{\,(W^{(k)} \otimes I\,)\,x^{(k)} \right. \\
 &-& \left. \frac{\alpha}{p_k} \left(\,\nabla F(x^{(k)}) \odot (z^{(k)}\otimes \mathbf{1})\,\right)\right\}, \nonumber
 \end{eqnarray}
where $W^{(k)} \otimes I$
denotes the Kronecker product of
$W^{(k)}$ and the $d \times d$ identity matrix, $\mathbf 1$ in~\eqref{eqn-update-compact} is of size $d \times 1$,
and $\odot$ denotes the Hadamard (entry-wise) product.
Note that sequence $\{x^{(k)}\}$ is a
sequence of random vectors, due to the randomness of the $z^{(k)}$'s.
 The case $p_k\equiv 1$, $\forall k$, corresponds to
standard distributed (sub)gradient method in~\cite{nedic_T-AC},
in which case~\eqref{eqn-update-compact} becomes:
 \begin{equation}
 \label{eqn-update-compact-nedic}
 x^{(k+1)} = \mathcal{P}_{\mathcal X^N}\left\{\,(\,C \otimes I\,)\,\,x^{(k)} - \alpha\,\nabla F(x^{(k)}) \right\}.
 \end{equation}

\section{Statement of main results}
\label{section-statement-main-results}
We now present our main results on the proposed
distributed method~\eqref{eqn-update-one-node}.
 For benchmarking of~\eqref{eqn-update-one-node},
 we first present a result on the convergence
 of standard distributed gradient algorithm~\eqref{eqn-update-compact-nedic}.
 All the results in the current section, together
 with some needed auxiliary results, are proved in Section~\ref{section-proofs}.
 Recall that $x^\star \in {\mathbb R}^d$
  is the solution to~\eqref{eqn-opt-prob-original}.
\begin{theorem}
   \label{theorem-standard-dis-grad-generic}
   Consider standard distributed gradient algorithm~\eqref{eqn-update-compact-nedic}
   with step-size $\alpha \leq \lambda_N(C)/L$. Then,
   $x^{(k)}$
    converges to a point $x^\bullet  = (\,(x_1^\bullet)^\top,...,(x_N^\bullet)^\top\,)^\top\in {\mathcal X}^{N}$ that satisfies, for all
    $i=1,...,N$:
    \begin{eqnarray}
    \label{eqn-Theorem-1-1}
    \|x^\bullet_i - x^\star \|^2 &\leq& \|x^\bullet - \mathbf 1 \otimes x^\star \|^2 \leq
    \alpha\, {\mathcal{C}_{\Psi}}\\
    {\mathcal{C}_{\Psi}} &:=& \frac{4 N (M_f-m_f)}{1-\lambda_2(C)} + \frac{2N G^2}{\mu\,(1-\lambda_2(C))}.
    \end{eqnarray}
    Furthermore:
   \begin{equation}
    \label{eqn-Theorem-1-2}
   \|x^{(k)} - x^\bullet \|  \leq 2\,\sqrt{N}\,D \,(1-\alpha\,\mu)^k= O \left( (1-\alpha\,\mu)^{k} \right).
   \end{equation}
   \end{theorem}
%
Theorem~\ref{theorem-standard-dis-grad-generic}
says that, with algorithm~\eqref{eqn-update-compact-nedic},
each node's estimate $x_i^{(k)}$
 converges to a point $x^\bullet_i$ in the neighborhood
 of the true solution $x^\star$;
 the distance of the limit $x^\bullet_i$
 from $x^\star$ is controlled by step-size $\alpha$ -- the smaller
  the step-size, the closer the limit to the true solution.
  Furthermore, $x_i^{(k)}$
  converges to a solution neighborhood (to $x_i^\bullet$)
   at a globally linear rate, equal to~$1-\alpha \mu$.
    Hence, there is a tradeoff with respect to
    the choice of $\alpha$: a small $\alpha$
     means a higher precision in the limit, but a
     slower rate to reach this precision. Note also
     that, for $\alpha \leq \lambda_N(C)/L$,
     the convergence factor $(1-\alpha\,\mu)$ does
     not depend on the underlying network, but the distance
     $\|x_i^\bullet - x^\star\|$ between arbitrary node $i$'s limit~$x_i^\bullet$
     and the solution~$x^\star$
      depends on the underlying network
      -- through the number of nodes $N$
       and the second largest eigenvalue of matrix~$C$.

\begin{remark}
\label{remark-1}
It is possible to extend Theorem~\ref{theorem-standard-dis-grad-generic}
 to allow also for the step-sizes
 $\alpha \in ( \lambda_N(C)/L,\,(1+\lambda_N(C))/L )$, in which case
 the convergence factor $(1-\alpha\,\mu)$ in~\eqref{eqn-Theorem-1-1}
  is replaced with $\max \left\{\alpha L-\lambda_N(C), 1-\alpha\,\mu \right\}$.
  We restrict ourselves to the case $\alpha \leq \lambda_N(C)/L$, both for simplicity
   and due to the fact that step-sizes $\alpha$  -- needed to
   achieve sufficient accuracies in practice --  are usually much smaller than $1/L$. (See also Section~\ref{section-simulations}.)
\end{remark}

\begin{remark}
\label{remark-2}
For $\alpha \leq \lambda_N(C)/L$, the convergence factor $(1-\alpha\,\mu)$
is an exact (tight) worst case convergence factor, in the following sense:
 given an arbitrary network and matrix $C$,
 and given an arbitrary step-size $\alpha \leq \lambda_N(C)/L$,
 there exists a specific choice of functions $f_i$'s, set $\mathcal X$,
 and initial point $x^{(0)} \in {\mathcal X}^N$,
 such that $\|x^{(k+1)}-x^\bullet\| = (1-\alpha\,\mu) \|x^{(k)}-x^\bullet\|$,
 for all $k=0,1,...$\footnote{Consider $f_i: \, {\mathbb R} \rightarrow {\mathbb R}$,
 $f_i(x) = x^2/2$, $\forall i$, $\mathcal X = \{x \in {\mathbb R}:\,|x| \leq 2\}$,
 and $x^{(0)}=1$. Note that, in this case, $x^\bullet = 0$, and $\mu=L=1$. For this example,
 it is easy to show that $\|x^{(k+1)}-x^\bullet\| = (1-\alpha) \|x^{(k)}-x^\bullet\| $,
 for all $k=0,1,...$, and so the convergence factor equals $1-\alpha\,\mu$.}
%
%
\end{remark}

  We benchmark the proposed method against the standard
   distributed gradient method by checking: 1) whether
   it converges to the same point $x^\bullet$; 2) if so,
   whether it converges linearly; and 3) if the convergence is linear,
   how the corresponding convergence factor compares with $(1-\alpha\,\mu)$ --
   the convergence factor of the standard distributed gradient method.

References~\cite{WotaoYinDisGrad,Matei}
 also analyze the convergence rate of the standard distributed gradient method,
 allowing for step-size ranges wider than $\alpha \in (0,\,\lambda_N(C)/L ]$.
 They establish bounds on quantity $\|x^{(k)} - \mathbf 1 \otimes x^\star \|$
 which are in general different than~\eqref{eqn-Theorem-1-2}, and
 they are not directly concerned with quantity $\|x^{(k)} - x^\bullet\|$, i.e.,
 precise characterization of convergence rate of $x^{(k)}$ towards its limit.
 We adopt here~\eqref{eqn-Theorem-1-2} as it gives an exact worst-case characterization
  of the convergence rate towards $x^\bullet$
  for $\alpha \in (0,\,\lambda_N(C)/L ]$ (see Remark~\ref{remark-2}).

We now state our main results on the proposed algorithm~\eqref{eqn-update-one-node}.
The first result deals with a more generic
sequence of the $p_k$'s that converge to one;
the second result is for the $p_k$'s
 that converge to one geometrically.
\begin{theorem}
\label{theorem-proposed-1-generic}
Consider algorithm~\eqref{eqn-update-one-node} with step-size
$\alpha \leq \lambda_N(C)/L$. Further, suppose that $p_k \geq p_{\mathrm{min}}$, $\forall k$,
for some $p_{\mathrm{min}}>0$, and let
$p_k \rightarrow 1$ as $k \rightarrow \infty$.
 Then, with algorithm~\eqref{eqn-update-one-node},
 the iterates $x^{(k)}$ converge, in the mean square sense,
 to the same point $x^\bullet$
   as the standard distributed gradient method~\eqref{eqn-update-compact-nedic},
   i.e., $\mathbb E \left[ \|x^{(k)} - x^\bullet\|^2 \right]
    \rightarrow 0$ as $k \rightarrow \infty$.
 Assume, in addition, that $p_k = 1-u_k$,
  with:
  \[
  0 \leq u_k \leq \frac{\mathcal{C}_u}{(k+1)^{1+\zeta}},\,\,\forall k,
  \]
  for some constants $\mathcal{C}_u >0$ and $\zeta>0$. Then,
  $x^{(k)}$ converges to $x^\bullet$ almost surely.
\end{theorem}

\begin{remark}
Note that Theorem~2 assumes that each individual~$f_i$ is strongly convex. An interesting
future research direction is to explore whether Theorem~2
can be extended such that one no longer requires that each~$f_i$ be convex, but that the aggregated cost~$f$
in~(1) is strongly convex.
\end{remark}

\begin{theorem}
\label{theorem-proposed-2-generic}
Consider algorithm~\eqref{eqn-update-one-node} with step-size
$\alpha \leq \lambda_N(C)/L$. Further, suppose that
$p_k  = 1 - \delta^{k+1}$, $k=0,1,...$,
for some $\delta \in (0,1)$,
and let $\eta: = \max\{1-\alpha \mu, \delta^{1/2}\}$. Then,
in the mean square sense, algorithm~\eqref{eqn-update-one-node}
  converges to the same point $x^\bullet$
   as the standard distributed gradient method~\eqref{eqn-update-compact-nedic},
   and, moreover:
\[
\mathbb E \left[ \|x^{(k)} - x^\bullet\|\right] = O\left( k \eta^k\right)
 = O\left( (\eta+\epsilon)^k\right),
\]
for arbitrarily small positive $\epsilon$. Furthermore,
if $\sqrt{\delta} \leq 1-\alpha \mu$:
\[
\mathbb E \left[ \|x^{(k)} - x^\bullet\|\right] = O\left( k (1-\alpha \mu)^k\right)
 = O\left( (1-\alpha \mu+\epsilon)^k\right).
\]
\end{theorem}
Theorem~\ref{theorem-proposed-1-generic}
states that, provided that the
$p_k$'s are uniformly bounded away from zero, from below,
and $p_k \rightarrow 1$, the method~\eqref{eqn-update-compact-nedic}
converges (in the mean square)
 to the same point as the standard distributed method~\eqref{eqn-update-compact-nedic}.
  If, moreover, $p_k$ converges to
  one at a sublinear rate at least $\frac{1}{(k+1)^{1+\zeta}}$
  (where $\zeta>0$ can be arbitrarily small), then
  the convergence also hold almost surely. Therefore, in such scenarios,
  the random idling schedule governed by the
 $p_k$'s does not affect the method's limit.

Theorem~\ref{theorem-proposed-2-generic} furthermore
suggests that, provided that convergence
of $p_k$ towards unity is linear (geometric) with
convergence factor ${\delta} \leq (1-\alpha \mu)^2$,
 algorithm~\eqref{eqn-update-one-node}~converges at the practically same rate as the standard method~\eqref{eqn-update-compact-nedic}, i.e.,
as if all the nodes were working all the time (albeit with
a larger hidden constant). Hence, we may expect that the proposed method~\eqref{eqn-update-one-node}
 achieves the same desired accuracy as~\eqref{eqn-update-compact-nedic} with
 lesser total cost (smaller number of
 the overall node activations--communications and computations).
  Still, this does not actually prove the reduction
  in total cost with the proposed method (due to
  ignoring hidden constants). However,
  we observed that the savings indeed occur in numerical implementations.
  Section~\ref{section-simulations} demonstrates this
   on logistic losses, but a similar behavior is observed on
   other costs like strongly convex quadratic and ``source localization'' costs (see,
   e.g.,~\cite{AsilomarNesterov} for the form of cost
   functions). The hidden convergence constant
   in Theorem~3 is dependent on the underlying network, sequence $\{p_k\}$,
   and step-size $\alpha$, and is given explicitly in Remark~\ref{remark-hidden-constant}.

We now provide an intuitive explanation as to why the proposed method~\eqref{eqn-update-one-node}
 reduces total cost with respect to the standard
 distributed gradient method~\eqref{eqn-update-compact-nedic}. To this end,
 introduce~$\overline{x}^{(k)}=\frac{1}{N}\sum_{i=1}^N x_i^{(k)}$ -- the
 (hypothetically available) global average of the nodes' estimates at iteration~$k$, and, for the sake of a clearer explanation, assume for a moment that problem~(1)
is unconstrained. Then, it is easy to verify that~$\overline{x}^{(k)}$ evolves according to the following recursion:
 \begin{equation}
 \label{eqn-intuition}
 \overline{x}^{(k+1)} = \overline{x}^{(k)} - \frac{\alpha}{p_k\,N}\sum_{i=1}^N \nabla f_i \left( x_i^{(k)}\right)\,z_i^{(k)},\,\,k=0,1,...
 \end{equation}
This is actually an inexact version of the (centralized)
hybrid stochastic-deterministic gradient method in~\cite{SchmidtHybrid}.
The ``inexactness'' here stems from the fact that
the term $ \nabla f_i \left( \overline{x}^{(k)}\right)$
 with the method in~\cite{SchmidtHybrid} is replaced
 with $ \nabla f_i \left( x_i^{(k)}\right)$ with~\eqref{eqn-update-one-node}.
 We will show ahead that, for all~$i$, $\mathbb E \left[\|x_i^{(k)} - \overline{x}^{(k)}\|\right] = O(\alpha)$,
 and hence the amount of ``inexactness'' with~\eqref{eqn-update-one-node} is
 relatively small. Therefore, the proposed
 distributed method~\eqref{eqn-update-one-node} behaves
 similarly as the centralized hybrid method in~\cite{SchmidtHybrid}.
  The authors of~\cite{SchmidtHybrid} interpret the method therein as a ``hybrid''
 of a stochastic gradient method and a standard gradient method;
 the initial algorithm phase (small $p_k$'s) corresponds
 to a stochastic gradient method, and the latter phase
 ($p_k$'s close to one) corresponds (nearly) to a standard gradient method.
 It is well known that stochastic gradient methods converge quickly in initial
 iterations but may be slow when near the solution, while having
 cheap iterations; standard gradient methods have a steady linear
 convergence rate and expensive iterations. Therefore, it is natural to
 consider a hybrid
 of the two -- stochastic gradient initially and standard gradient at a later stage.
 This is why the method in~\cite{SchmidtHybrid} saves computations
 with respect to the standard (centralized) gradient method. For similar reasons,
 our proposed distributed method~\eqref{eqn-update-one-node} saves cost over the
 standard distributed gradient method~\eqref{eqn-update-compact-nedic}.

\textbf{Choice of~$p_k$}. We now comment on the choice of
$p_k$ and $\delta$. Note from Theorem~3
that $\delta=(1-\alpha\mu)^2$ is the
largest, i.e., ``most aggressive,''
value of $\delta$ which guarantees the best possible convergence factor
 (which equals to $1-\alpha \mu + \epsilon$, $\epsilon>0$ arbitrarily small).
 Hence, it is the choice which
 spends the lowest cost while guaranteeing
 the best possible convergence factor, and hence it is
 optimal in the above sense. For practical implementations,
 when problems under study are very ill-conditioned (
 with the very large value of $L/\mu$),
 we recommend the following modification:
  $p_k = \max\{1-\delta^{k+1},\,\underline{p}\}$,
 $k=0,1,...,$ and $\delta =\min\{(1-\alpha\,\mu)^2,\,\overline{\delta}\},$
 $\overline{\delta}>0$. Quantity~$\underline{p}$
   prevents using very small values of $p_k$
    in the initial iterations,
    while~$\overline{\delta}$
 prevents too slow increase of $p_k$
towards unity for very ill-conditioned problems and
very small step-sizes. A possible choice
is $\underline{p}=0.1,$ and~$\overline{\delta}=0.99999.$

\section{Intermediate results and proofs}
\label{section-proofs}
Subsection~\ref{subsection-disagreements}
gives intermediate results on
the random matrices $W^{(k)}$
 and provides the disagreement estimates --
 how far apart are the estimates $x_i^{(k)}$
  of different nodes in the network.
   Subsection~\ref{subsection-penalty-analysis}  introduces a
    penalty-like interpretation of algorithm~\eqref{eqn-update-compact-nedic}
     and proves Theorem~\ref{theorem-standard-dis-grad-generic}.
     Finally, Subsection~\ref{subsection-proposed-analysis}
     proves our main results, Theorems~\ref{theorem-proposed-1-generic} and~\ref{theorem-proposed-2-generic},
      by applying the penalty-like interpretation on algorithm~\eqref{eqn-update-compact}.
 For notational simplicity, this section presents
 auxiliary results and all the proofs for the case
 $d=1$, but all these extend to a generic~$d>1$. Throughout this Section,
all the claims (equalities and inequalities) which deal with random quantities hold either:
1) surely, for any random realization; or 2) in expectation. It is clear from notation which of the two cases is in force.

\subsection{Matrices~$W^{(k)}$ and disagreement estimates}
\label{subsection-disagreements}
\textbf{Matrices~$W^{(k)}$}.
Recall that $J : = (1/N) \mathbf 1 \mathbf 1^\top$. We have the following Lemma
on the matrices $W^{(k)}$. Lemma~\ref{lemma-matrices-W-k}
 follows from simple arguments and
 standard results on symmetric, stochastic matrices
  (see, e.g.,~\cite{bajovicProducts}).
 Hence, we omit the proof for brevity.
%
%
\begin{lemma}[Matrices $W^{(k)}$]
\label{lemma-matrices-W-k}
\begin{enumerate}[(a)]
\item The sequence $\{W^{(k)}\}$ is a sequence of independent random matrices.
\item For all $k$, $W^{(k)}$ is symmetric and stochastic (rows sum to one
and all the entries are nonnegative).
\item For all $k$, $0 \prec W^{(k)} \preceq I.$
\item There exists a constant $\beta \in (0,1)$ such that, $\forall k$,
 $\mathbb E \left[ \|W^{(k)}-J\|^2 \right] < \beta^2$.
\end{enumerate}
\end{lemma}

It can be shown that
$\beta$ can be taken as
 $\beta^2 = 1-(p_{\mathrm{min}})^N \left[\, 1-(\lambda_2(C))^2\,\right];$
 see, e.g.,~\cite{bajovicProducts}.
%

\begin{remark}
The quantities~$\mathbb E \left[ \|W^{(k)}-J\|^2 \right]$
clearly depend on~$k$, and, more specifically, on $p_k$.
 We adopt here a (possibly loose) uniform bound~$\beta$
 (independent of~$k$) as
this suffices to establish conclusions about
convergence rates of algorithm~\eqref{eqn-update-compact}
 while simplifying the presentation.
\end{remark}

\textbf{Disagreement estimate}. Recall
 $\overline{x}^{(k)}:=\frac{1}{N}\sum_{i=1}^N x_i^{(k)}$ --
  the global average of the nodes' estimates,
  and denote by $\widetilde{x}_i^{(k)}=x_i^{(k)} - \overline{x}^{(k)}$.
   Note that both quantities are random.
  The quantity $\widetilde{x}_i^{(k)}$ measures how far is
  the node $i$'s estimate from the global average.
  Denote by $\widetilde{x}^{(k)}:=(\widetilde{x}_1^{(k)},...,\widetilde{x}_N^{(k)})^\top$.
  The next Lemma shows that $\mathbb E \left[\|\widetilde{x}^{(k)}\|^2\right]$ is
  uniformly bounded, $\forall k$, and that the bound is $O(\alpha^2)$, i.e.,
  the disagreement size is controlled by the step-size. (The smaller
  the step-size, the smaller the disagreements are.)
  The proof follows similar arguments as, e.g.,~\cite{nedic_T-AC-private},
  and it can be found in the Appendix.
\begin{lemma}[Disagreements bound]
\label{lemma-disagreements}
For all $k$, there holds:
\[
\mathbb E \left[\|\widetilde{x}^{(k)}\|^2\right] \leq \left(\frac{3 \alpha \sqrt{N} G}{p_{\mathrm{min}} (1-\beta)}\right)^2.
\]
\end{lemma}

\subsection{Analysis of the standard distributed gradient method through a penalty-like reformulation}
\label{subsection-penalty-analysis}
We analyze the proposed method~\eqref{eqn-update-compact}
 through a penalty-like interpretation,
 to our best knowledge first introduced in~\cite{cdc-submitted}.
Introduce an auxiliary function
 $\Psi_{\alpha}: \, {\mathbb R}^N \mapsto \mathbb R$, defined by:
  $\Psi_{\alpha}(x):=\sum_{i=1}^N f_i(x_i)
   + \frac{1}{2 \alpha} x^\top (I-C) x $
    $=F(x)
   + \frac{1}{2 \alpha} x^\top (I-C) x$, and the
   associated optimization problem:
   \begin{equation}
\label{eqn-opt-prob-penalty}
\begin{array}[+]{ll}
\mbox{minimize} \:\:\Psi_{\alpha}(x)=\sum_{i=1}^N f_i(x_i)
   + \frac{1}{2 \alpha} x^\top (I-C) x\\
\mbox{subject to}\:\: x \in \mathcal{X}^N.
\end{array}
\end{equation}
Function $\Psi_{\alpha}$ and~\eqref{eqn-opt-prob-penalty}
 will be very useful in the analysis of~\eqref{eqn-update-one-node}.
   In fact, we will show that
   \eqref{eqn-update-one-node} is an inexact version of the (projected) gradient
   method on function~$\Psi_{\alpha}$.
   Clearly, \eqref{eqn-opt-prob-penalty} is solvable, and it has a
   unique solution, which we denote by~$x^{\bullet}$.\footnote{The point of convergence
 of algorithm~\eqref{eqn-update-compact-nedic}
 and the solution to~\eqref{eqn-opt-prob-penalty}
 are intentionally denoted by the same
 symbol because -- as we will show -- they actually are the same point.}

   We start by showing
   that standard distributed (sub)gradient method in~\cite{nedic_T-AC}
   is an exact (projected) gradient method on~$\Psi_{\alpha}$.
    Indeed, the derivative
   $\nabla \Psi_{\alpha}(x) = \nabla F(x) + \frac{1}{\alpha}(I-C)x$.
   The projected gradient method on~$\Psi_{\alpha}$ with
   step-size $\alpha$ then takes the form:
   \begin{eqnarray}
   \label{eqn-update-penalty}
   x^{(k+1)} &=&
     \mathcal{P}_{\mathcal X^N} \left\{x^{(k)} - \alpha \nabla \Psi_{\alpha}(x^{(k)})\right\}\\
     &=&
    \mathcal{P}_{\mathcal X^N} \left\{ x^{(k)}  - \right. \nonumber \\
    &\,& \left. \alpha \left( \nabla F(x^{(k)}) + \frac{1}{\alpha}(I-C)x^{(k)} \right)\right\}, \nonumber
   \end{eqnarray}
   which, after rearranging terms, is precisely~\eqref{eqn-update-compact-nedic}.

 It is easy to see that $\Psi_{\alpha}$ is strongly convex on ${\mathbb R}^N$,
 with modulus $\mu^\prime = \mu$ (which equals the
 strong convexity modulus of the $f_i$'s).
   Further, $\nabla \Psi_{\alpha}$ is
    Lipschitz continuous on ${\mathbb R}^N$, with constant
     $L^\prime
     = L + \frac{1-\lambda_N(C)}{\alpha}$.
     Namely, $\forall x,y \in {\mathbb R}^N$:
      $
     \|\nabla \Psi_{\alpha}(x) - \nabla \Psi_{\alpha}(y) \|
     $ $\leq$ $
     \|\nabla F(x) - \nabla F(y)\| $ $
     $ $+$ $\frac{1}{\alpha} \|I-C\|\,\|x-y\|$ $\leq$
    $ L \|x-y\| + \frac{1-\lambda_N(C)}{\alpha} \|x-y\|.
      $
     %
     (Note that $\|\nabla F(x) - \nabla F(y)\| \leq L \|x-y\|$ follows
     after summing the inequalities:
     $|\nabla f_i(x_i) - \nabla f_i(y_i)|^2 \leq L^2 |x_i-y_i|^2$, $i=1,...,N$,
     and using $\| \nabla F(x)\|^2 = \sum_{i=1}^N |\nabla f_i(x_i)|^2$.)
      We impose that $\alpha$ satisfies $\alpha \leq \frac{1}{L^\prime}$,
      which, after simple manipulations,
      gives: $\alpha \leq \lambda_N(C)/(L)$, as introduced before.

   An immediate consequence of
   the fact that algorithm~\eqref{eqn-update-compact-nedic}
   is precisely the projected gradient method
   to solve~\eqref{eqn-opt-prob-penalty}
   is the following Lemma, first observed in~\cite{cdc-submitted}.

   \begin{lemma}[\cite{cdc-submitted}]
   \label{lemma-cdc}
   Standard distributed gradient algorithm~\eqref{eqn-update-compact-nedic}
   with step-size $\alpha \leq \lambda_N(C)/L$ converges
   to the point~$x^\bullet \in {\mathcal X}^N$ -- the solution to~\eqref{eqn-opt-prob-penalty}.
   \end{lemma}

We proceed by proving Theorem~\ref{theorem-standard-dis-grad-generic}.
   \begin{IEEEproof}[Proof of Theorem~\ref{theorem-standard-dis-grad-generic}]
   As per Lemma~\ref{lemma-cdc},
   algorithm~\eqref{eqn-update-compact-nedic}
    converges to $x^\bullet$ -- the solution to~\eqref{eqn-opt-prob-penalty}.
    We hence need to prove for the solution to~\eqref{eqn-opt-prob-penalty}
     the characterization in~\eqref{eqn-Theorem-1-1}.

   Consider an arbitrary point $x \in {\mathcal X}^N$,
   and let $\overline{x}:=\frac{1}{N}\sum_{i=1}^N x_i$.
   We first prove the following inequality:
   \begin{equation}
   \label{eqn-proof-aux-new}
   f(\overline{x})-f(x^\star) \leq \left(\Psi_{\alpha}(x)
   - \Psi_{\alpha}(x^\bullet)\right) + \frac{\alpha N G^2}{2(1-\lambda_2(C))}.
   \end{equation}
Indeed, we have that:
   \begin{eqnarray*}
   x^\top(I-C)x  &=& (x-\overline{x} \mathbf 1)^\top(I-C) (x-\overline{x} \mathbf 1) \\
   &\geq& \lambda_{N-1}(I-C)\|x-\overline{x} \mathbf 1\|^2\\
     &= &(1-\lambda_2(C)) \|\widetilde{x}\|^2,
     \end{eqnarray*}
     where we let $\widetilde{x}:=x-\overline{x} \mathbf 1$.
      Further,
      {\allowdisplaybreaks{
      \begin{eqnarray*}
      \sum_{i=1}^N f_i(x_i)
       &=& \sum_{i=1} f_i(\overline{x}) +
       (\,\sum_{i=1} (f_i(x_i) - f_i(\overline{x}))\,) \\
        &\geq& f(\overline{x})
         - G \sum_{i=1}^N |x_i - \overline{x}|\\
          &\geq& f(\overline{x})
         - G \sqrt{N} \|\widetilde{x}\|.
         \end{eqnarray*}}}
         The second from last inequality follows because
         $f_i(x_i)  \geq f_i(\overline{x}) + \nabla f_i(\overline{x})(x_i-\overline{x})
          \geq f_i(\overline{x}) -G \,|x_i-\overline{x}|$.
           Combining the previous conclusions:
           {\allowdisplaybreaks{
           \begin{eqnarray}
           \Psi_{\alpha}(x) - \Psi_{\alpha}(x^\bullet)
           &\geq&
           f(\overline{x}) - \Psi_{\alpha}(x^\bullet) - G \sqrt{N} \|\widetilde{x}\| \nonumber\\
           &+& \frac{1}{2\,\alpha} (1-\lambda_2(C)) \|\widetilde{x}\|^2 \nonumber \\
           &\geq&
           f(\overline{x}) - \Psi_{\alpha}(x^\bullet) \nonumber\\
           &-& \sup_{t \geq 0}
           \left\{G \sqrt{N} t - \frac{1}{2 \alpha}(1-\lambda_2(C)) t^2\right\} \nonumber \\
           &\geq&
           \label{eqn-chain-new-1}
           f(\overline{x}) \hspace{-1mm}-\hspace{-1mm} \Psi_{\alpha}(x^\bullet) - \frac{\alpha\,N\,G^2}{2(1-\lambda_2(C))}.
           \end{eqnarray}}}
         Next, note that
         $\Psi_{\alpha}(x^\bullet) =
         \min_{x \in \mathcal{X}^N}\Psi_{\alpha}(x)
         \leq \Psi_{\alpha}(x^\star \mathbf 1) = f(x^\star)$,
         and so $-\Psi_{\alpha}(x^\bullet)\geq - f(x^\star).$ Applying this
         to~\eqref{eqn-chain-new-1},
          completes the proof of~\eqref{eqn-proof-aux-new}.

          We now prove claim~\eqref{eqn-Theorem-1-1} in Theorem~\ref{theorem-standard-dis-grad-generic}.
          We have:
          \begin{eqnarray}
          \left\| x^\bullet - x^\star \mathbf 1 \right\|^2
          &=&
          \left\| x^\bullet - \overline{x}^\bullet \mathbf 1 + \overline{x}^\bullet \mathbf 1 - x^\star \mathbf 1 \right\|^2
          \nonumber \\
          &\leq&
          \label{eqn-proof-part-b}
          2 \left\| x^\bullet - \overline{x}^\bullet \mathbf 1 \right\|^2
          + 2 N  \left|\overline{x}^\bullet- x^\star  \right|^2.
          \end{eqnarray}
          For the second summand in~\eqref{eqn-proof-part-b}, we have:
          \begin{eqnarray*}
          \left|\overline{x}^\bullet- x^\star  \right|^2 &\leq& \frac{2}{N \mu} \left(
          f(\overline{x}^\bullet)- f(x^\star)\right) \\
          &\leq&
          \frac{\alpha N G^2}{N\mu(1-\lambda_2(C))}.
          \end{eqnarray*}
          The first inequality above is
          due to strong convexity of $f$ (with modulus~$N \mu$),
          and the second applies \eqref{eqn-proof-aux-new}
          with $x = x^\bullet$ (where $\overline{x}^\bullet = \frac{1}{N}\sum_{i=1}^N x^\bullet_i$).
            We now upper bound the first summand in~\eqref{eqn-proof-part-b}.
            We have that:
            \begin{eqnarray*}
            \Psi_{\alpha}(x^\bullet) &=& \sum_{i=1}^N f_i(x_i^\bullet) + \frac{1}{2 \alpha} (x^\bullet)^\top (I-C) x^\bullet \\
            &\geq& \frac{1-\lambda_2(C)}{2 \alpha}
            \|\widetilde{x}^\bullet\|^2  + N m_f,
            \end{eqnarray*}
           where $\widetilde{x}^\bullet = x^\bullet - \overline{x}^\bullet \,\mathbf 1$.
          On the other hand,
          \[
            \Psi_{\alpha}(x^\bullet) \leq f(x^\star) \leq N \,M_f.
            \]
          Combining the obtained upper and lower bounds on $\Psi_{\alpha}(x^\bullet) $,
          we obtain for the first summand in~\eqref{eqn-proof-part-b}:
          \[
          \left\| x^\bullet - \overline{x}^\bullet \mathbf 1 \right\|^2 \leq \frac{2 \alpha N (M_f - m_f)}{1-\lambda_2(C)}.
          \]
          Combining the bounds on the first and second summands, the claim in \eqref{eqn-Theorem-1-1} follows.%
%


It remains to prove the claim in~\eqref{eqn-Theorem-1-2}.
 By standard analysis of gradient methods, we have that:
\[
\|x^{(k)} - x^\bullet \| \leq (1-\alpha \mu)^{k}
\|x^{(0)} - x^\bullet \| \leq (1-\alpha \mu)^{k} 2 \sqrt{N} D,
\]
 where we used that $\|x^{(0)}\| \leq \sqrt{N} D$, and the same bound for $x^\bullet$.
   Thus, the desired result.
 \end{IEEEproof}

 \subsection{Analysis of the proposed method~\eqref{eqn-update-one-node}}
 \label{subsection-proposed-analysis}
   We now turn our attention to
   the proposed method~\eqref{eqn-update-compact}.
   It is easy to verify that~\eqref{eqn-update-compact}
    can be written as:
    \begin{eqnarray}
    \label{eqn-proposed-analysis-form}
   x^{(k+1)} =
     \mathcal{P}_{\mathcal X^N} \left\{x^{(k)} - \alpha \left[\nabla \Psi_{\alpha}(x^{(k)}) + e^{(k)}\right]\right\},
     \end{eqnarray}
   where $e^{(k)} = (e_1^{(k)},...,e_N^{(k)})^\top$
    is a random vector, with $i$-th component equal to:
    {\allowdisplaybreaks{
    \begin{eqnarray}
    e_i^{(k)} &=& \left( \frac{z_i^{(k)}}{p_k}-1\right)\nabla f_i(x_i^{(k)}) \nonumber \\
    \label{eqn-e-i}
    &+&    \frac{1}{\alpha}\sum_{j \in \Omega_i } C_{ij} \,(z_i^{(k)} z_j^{(k)}-1) \left( x_i^{(k)} - x_j^{(k)}\right).
    \end{eqnarray}}}
    Hence,~\eqref{eqn-update-one-node} is an inexact projected
    gradient method applied to $\Psi_{\alpha}$, with step-size
    $\alpha$, where the amount of inexactness is given by vector~$e^{(k)}$.

Overall, our strategy in analyzing~\eqref{eqn-proposed-analysis-form}
consists of two main steps: 1) analyzing
the inexact projected gradient method \eqref{eqn-proposed-analysis-form}; and 2)
characterizing (upper bounding) the inexactness vector~$e^{(k)}$.
 For the former step, we apply Proposition~3 in~\cite{Schmidt}.
  Adapted to our setting, the proposition says the following.
  Consider minimization of $\phi(y)$ over $y \in \mathcal Y$,
  where $\phi: \, \mathbb R^m \rightarrow \mathbb R$ is a
  convex function, and $\mathcal Y \subset \mathbb R^m$ is a closed convex set.
  Let $y^\bullet$ be the solution to the above problem. Further, let $\phi$ be strongly convex with modulus $\mu_{\phi}>0$,
  and let $\phi$ have a Lipschitz continuous gradient with constant~$L_{\phi} \geq \mu_{\phi}$.

  \begin{lemma}[Proposition~3, \cite{Schmidt}]
  Consider the algorithm:
  \[
  y^{(k+1)} = \mathcal{P}_{\mathcal Y}\left\{ y^{(k)} - \frac{1}{L_{\phi}} \left[ \nabla \phi(y^{(k)}) + e_{y}^{(k)}\right]\right\}
  ,\,\,
  k=0,1,...,\]
  where $e_{y}^{(k)}$ is a random vector.
  Then, $\forall k=1,2,...$:
  \begin{eqnarray}
\|y^{(k)}-y^\bullet\| &\leq& (1-\mu_{\phi}/L_{\phi})^k \|y^{(0)} - y^\bullet\| \nonumber \\
\label{eqn-Schmidt-1}
&+& \frac{1}{L_{\phi}}
\sum_{t=1}^k (1-\mu_{\phi}/L_{\phi})^{k-t} \|e_y^{(t-1)}\|,
\end{eqnarray}
where $y^{(0)} \in \mathcal Y$ is the initial point.
\end{lemma}
Note that, if $\nabla \phi$ is Lipschitz continuous
with constant $L_{\phi}$, then $\nabla \phi$ is also Lipschitz
 continuous with constant $1/\alpha \geq L_{\phi}$.
 Therefore, for the function $\phi$
  and the iterations:
  \[
  y^{(k+1)} = \mathcal{P}_{\mathcal Y}\left\{ y^{(k)} - \alpha
  \left[ \nabla \phi(y^{(k)}) + e_{y}^{(k)}\right]\right\},
  \,\,
k=0,1,...,
\]
 there holds:
\begin{eqnarray}
\|y^{(k)}-y^\bullet\| &\leq& (1-\alpha\,\mu_{\phi})^k \|y^{(0)} - y^\bullet\| \nonumber \\
\label{eqn-Schmidt-2}
&+& \hspace{-2mm}\alpha
\sum_{t=1}^k (1-\alpha\,\mu_{\phi})^{k-t} \|e_y^{(t-1)}\|, \,k=1,...
\end{eqnarray}
In other words, the modified claim~\eqref{eqn-Schmidt-2} holds
even if we take a step size different (smaller than)~$1/L_{\phi}$.

For analyzing the inexact projected
 gradient method~\eqref{eqn-proposed-analysis-form}, we will also make use
 of the following result. Claim~(17) is Lemma 3.1 in in~\cite{nedic_novo}.
  Claim~(18) can be proved by following similar arguments as in equations~(44)-(48) in~\cite{arxivVersion}.
\begin{lemma}
\label{lemma-for-proof-of-thm-2}
Consider a
deterministic sequence $\{v_k\}_{k=0}^{\infty}$ such that $v_k \rightarrow 0$ as $k \rightarrow \infty$, and
let $a$ be a constant in $(0,1)$. Then, there holds:
\begin{equation}
\label{eqn-lemma-for-proof-of-thm-2}
\sum_{t=1}^k a^{k-t} v_{t-1} \rightarrow 0.
\end{equation}
If, moreover, there exist positive constants $\mathcal C_v$
 and $\zeta$ such that, for all $k=0,1,...$,
 \[
 0 \leq v_k \leq \frac{\mathcal C_v}{(k+1)^{1+\zeta}},
 \]
then there exists positive constant $\mathcal C_v^\prime $ such that, for all $k=1,2,...$,
\begin{equation}
\label{eqn-lemma-for-proof-of-thm-2-new}
\sum_{t=1}^k a^{k-t} v_{t-1} \leq \frac{\mathcal C_v^\prime}{k^{1+\zeta}}.
\end{equation}
\end{lemma}

\textbf{Step 1: gradient inexactness}. We proceed by characterizing the gradient inexactness;
 Lemma~\ref{lemma-gradient-inexactness} upper bounds quantity~$\mathbb E \left[ \|e^{(k)}\|^2 \right]$.
     \begin{lemma}[Gradient inexactness]
     \label{lemma-gradient-inexactness}
     For all $k=0,1,...$, there holds:
     \begin{eqnarray}
     \mathbb E \left[ \|e^{(k)}\|^2 \right] &\leq& 4 (1-p_k) \frac{N \,G^2}{p_{\mathrm{min}}} \nonumber\\
     &+& 72 (1-p_k^2) \frac{N G^2}{(p_{\mathrm{min}})^2 (1-\beta)^2} \nonumber \\
     &\leq&  {\mathcal{C}_{e}} \,(1-p_k^2),
     \end{eqnarray}
     where
     \begin{eqnarray}
     \label{eqn-C-e-value}
     {\mathcal{C}_{e}} = \frac{4\,N \,G^2}{p_{\mathrm{min}}}
     + \frac{72\, N G^2}{(p_{\mathrm{min}})^2 (1-\beta)^2}.
     \end{eqnarray}
     \end{lemma}
 \begin{IEEEproof}
 Consider~\eqref{eqn-e-i}. We have:
  \begin{eqnarray}
    \label{eqn-chain-ineq-1}
    |e_i^{(k)}|^2 &\leq& 2\,\left| \frac{z_i^{(k)}}{p_k}-1\right|^2 \,|\nabla f_i(x_i^{(k)})|^2 \\
    &+&
    \frac{2}{\alpha^2}\,\sum_{j \in \Omega_i } C_{ij} \,|z_i^{(k)} z_j^{(k)}-1|^2\, \left| x_i^{(k)} - x_j^{(k)}\right|^2 \nonumber \\
    \label{eqn-chain-ineq-2}
    &\leq&
    2 G^2\,\left| \frac{z_i^{(k)}}{p_k}-1\right|^2  + \frac{4}{\alpha^2}\,
    \sum_{j \in \Omega_i } C_{ij} \,|z_i^{(k)} z_j^{(k)}-1|^2 \nonumber \\
    &\times& \left(\left| \widetilde{x}_i^{(k)}
    \right|^2 + \left| \widetilde{x}_j^{(k)}\right|^2\right).
    \end{eqnarray}
    Inequality~\eqref{eqn-chain-ineq-1} uses
     the following bound: $(u+v)^2 \leq 2 u^2 + 2 v^2$.
      It also
       uses, with
       $u_i:=(z_i^{(k)} z_j^{(k)}-1)( x_i^{(k)} - x_j^{(k)})$,
       the following relation:
       \begin{eqnarray*}
       (\sum_{j \in \Omega_i}C_{ij}u_j)^2
        &=&
        (\sum_{j \in \Omega_i}C_{ij}u_j + C_{ii}\cdot 0)^2 \\
         &\leq& \sum_{j \in \Omega_i}C_{ij}u_j^2 + C_{ii}\cdot 0^2\\
          &=&\sum_{j \in \Omega_i}C_{ij}u_j^2,
          \end{eqnarray*}
          which follows due to the fact that
          $\sum_{j \in \Omega_i}C_{ij}u_j + C_{ii}\cdot 0$
           is a convex combination, and
           $v \mapsto v^2$, $v \in \mathbb R$, is convex.
            Inequality~\eqref{eqn-chain-ineq-2}
            uses that
            \begin{eqnarray*}
            \left| x_i^{(k)} - x_j^{(k)}\right|^2  &=&
            \left| x_i^{(k)} - \overline{x}^{(k)} + \overline{x}^{(k)} - x_j^{(k)}\right|^2 \\
             &\leq&
             2 \left| x_i^{(k)} - \overline{x}^{(k)}\right|^2 +
             2\left|\overline{x}^{(k)} - x_j^{(k)}\right|^2.
             \end{eqnarray*}
      Taking expectation,
     and using independence of~$x^{(k)}$ from~$z^{(k)}$:
     {\allowdisplaybreaks{
    \begin{eqnarray}
    \mathbb E \left[ |e_i^{(k)}|^2 \right]
    &\leq&
    2 G^2\,\mathbb E \left[\left| \frac{z_i^{(k)}}{p_k}-1\right|^2 \right] \nonumber \\
    &+& \frac{4}{\alpha^2}\,
    \sum_{j \in \Omega_i } C_{ij} \,\mathbb E \left[|z_i^{(k)} z_j^{(k)}-1|^2\right] \nonumber \\
    &\times& \left(\mathbb E \left[\left| \widetilde{x}_i^{(k)}
    \right|^2 \right]
    \label{eqn-expectations-main}
    + \mathbb E \left[\left| \widetilde{x}_j^{(k)}\right|^2\right]\right).
    \end{eqnarray}}}
    We proceed by upper bounding $\mathbb E \left[\left| \frac{z_i^{(k)}}{p_k}-1\right|^2 \right]$,
    using the total probability law with respect to the following partition:
    $\{z_i^{(k)}=1\}$, and $\{z_i^{(k)}=0\}$:
    \begin{eqnarray}
    &\,& \hspace{-8mm}\mathbb E \left[ \left| \frac{z_i^{(k)}}{p_k}-1\right|^2 \right]
     =
    \left| \frac{1}{p_k}-1\right|^2 \mathbb P(z_i^{(k)}=1) + P\left(z_i^{(k)} = 0\right)
    \nonumber \\
    & = &
    \left| \frac{1}{p_k}-1\right|^2 p_k + (1-p_k)\\
    & =&
    \frac{1}{p_k}(1-p_k)^2 + (1-p_k)  \nonumber  \\
    &\leq&
     \frac{1}{p_k}(1-p_k) + (1-p_k) \nonumber \\
     \label{eqn-expectation-aux-1}
     &\leq& 2(1-p_k)/p_{\mathrm{min}}.
    \end{eqnarray}
    We next upper bound $\mathbb E \left[|z_i^{(k)} z_j^{(k)}-1|^2\right]$,
    using the total probability law with respect to
    the event $\{z_i^{(k)}=1,\,z_j^{(k)}=1\}$ and its complement; we obtain:
    \begin{eqnarray}
    \mathbb E \left[ |z_i^{(k)} z_j^{(k)}-1|^2 \right]
     &=&
     (1-\mathbb P(z_i^{(k)}=1,\,z_j^{(k)}=1)) \nonumber\\
    \label{eqn-expectation-aux-2}
    &=&
    (1-p_k)^2 .
    \end{eqnarray}
Substituting~\eqref{eqn-expectation-aux-1} and~\eqref{eqn-expectation-aux-2} in~\eqref{eqn-expectations-main}:
{\allowdisplaybreaks{
\begin{eqnarray}
    \mathbb E \left[ |e_i^{(k)}|^2 \right]
    &\leq&
    4\,G^2\,(1-p_k)/p_{\mathrm{min}} \nonumber \\
    &+&
    \frac{4}{\alpha^2}\,\sum_{j \in \Omega_i } C_{ij} \,(1-p_k^2) \nonumber\\
    &\times&
    \, \left(\mathbb E \left[\left| \widetilde{x}_i^{(k)}
    \right|^2 \right]
    +
    \label{eqn-expectations-main-2}
    \mathbb E \left[\left| \widetilde{x}_j^{(k)}\right|^2\right]\right).
    \end{eqnarray}}}
Summing the above inequalities over $i=1,...,N$, using
the fact that $\sum_{j \in \Omega_i} C_{ij} \leq 1$, $\forall i$,
 $\mathbb E \left[ \|e^{(k)}\|^2 \right] =
\sum_{i=1}^N \mathbb E \left[ |e_i^{(k)}|^2 \right]$,
and
$\mathbb E \left[ \|{\widetilde x}^{(k)}\|^2 \right] =
\sum_{i=1}^N \mathbb E \left[ |{\widetilde x}_i^{(k)}|^2 \right]$,
  we obtain:
 \begin{eqnarray*}
    \mathbb E \left[ \|e^{(k)}\|^2 \right]
    &\leq&
    4\,N\,G^2\,(1-p_k)/p_{\mathrm{min}} \\
    &+&
    \frac{8}{\alpha^2}\, (1-p_k^2)
    \, \mathbb E \left[\left\| \widetilde{x}^{(k)}
    \right\|^2 \right] .
    \end{eqnarray*}
    Finally, applying Lemma~\ref{lemma-disagreements} to the last inequality,
    the claim follows.
 \end{IEEEproof}

\textbf{Step 2: Analyzing the inexact projected gradient method}. We first state and
 prove the following Lemma on algorithm~\eqref{eqn-update-one-node}.
\begin{lemma}
\label{theorem-proposed-alg}
Consider algorithm~\eqref{eqn-update-one-node} with step-size
$\alpha \leq \lambda_N(C)/(L)$. Then, for the iterates $x^{(k)}$
 and $x^\bullet$--the solution to~\eqref{eqn-opt-prob-penalty}, $\forall k=1,2,...,$ there holds:
 \begin{eqnarray*}
 \mathbb E \left[ \|x^{(k)}-x^\bullet\|^2 \right]
 &\leq& 8\,N\, (1-\alpha \mu )^{2 k} D^2 \\
 &+&
 \frac{\alpha\,{\mathcal{C}_{e}}}{\mu} \sum_{t=1}^k (1-\alpha \,\mu)^{k-t}(1-p_{t-1}^2).
 \end{eqnarray*}
\end{lemma}
\begin{IEEEproof}
As already established, algorithm~\eqref{eqn-update-one-node}
is an inexact projected gradient method to solve~\eqref{eqn-opt-prob-penalty},
with the inexactness vector $e^{(k)}$.
We now apply~\eqref{eqn-Schmidt-2} to sequence $x^{(k)}$
 and iterations~\eqref{eqn-update-compact}; we obtain:

%
%
\begin{eqnarray}
\|x^{(k)}-x^\bullet\| &\leq& (1-\alpha \,\mu)^k \|x^{(0)} - x^\bullet\| \nonumber \\
\label{eqn-to-take-expectation}
&+& \alpha \sum_{t=1}^k (1-\alpha \mu)^{k-t} \|e^{(t-1)}\|.
\end{eqnarray}
Squaring the latter inequality, using $(u+v)^2 \leq 2 u^2 + 2 v^2$,
and $\|x^{(0)} - x^\bullet\|\leq 2 \sqrt{N} D$:
\begin{eqnarray}
\|x^{(k)}-x^\bullet\|^2 &\leq& 8(1-\alpha \,\mu)^{2k} N D^2 \nonumber \\
&+& \alpha^2 \left( \sum_{t=0}^k (1-\alpha \mu)^{k-t}\right) \nonumber \\
\label{eqn-aux-new}
&\times&
\sum_{t=1}^k (1-\alpha \mu)^{k-t} \|e^{(t-1)}\|^2.
\end{eqnarray}
In \eqref{eqn-aux-new},
we used the following. Let $\theta_t = (1-\alpha \mu)^{k-t} $,
and $S_t: = \sum_{t=1}^k \theta_t$.
Then,
\begin{eqnarray*}
\left(\sum_{t=1}^k \theta_t \|e^{(t-1)}\|\right)^2
&=&
S_t^2 \left(\sum_{t=1}^k \frac{\theta_t}{S_t} \|e^{(t-1)}\|\right)^2 \\
  &\leq &
  S_t^2
  \sum_{t=1}^k \frac{\theta_t}{S_t} \|e^{(t-1)}\|^2\\
  &=& S_t
  \sum_{t=1}^k {\theta_t} \|e^{(t-1)}\|^2,
  \end{eqnarray*}
   where we used convexity of the scalar quadratic function $v \mapsto v^2$.
Now, using $\sum_{t=1}^k (1-\alpha \mu)^{k-t} \leq \frac{1}{1-(1-\alpha \mu)} = \frac{1}{\alpha \mu}$,
\eqref{eqn-aux-new} is further upper bounded as:
\begin{eqnarray}
\|x^{(k)}-x^\bullet\|^2 &\leq& 8(1-\alpha \,\mu)^{2k} N D^2 \nonumber \\
&+& \frac{\alpha^2}{\alpha \mu}
\sum_{t=1}^k (1-\alpha \mu)^{k-t} \|e^{(t-1)}\|^2. \nonumber
\end{eqnarray}
Taking expectation, and applying Lemma~\ref{lemma-gradient-inexactness}, we obtain the claimed result.
\end{IEEEproof}

We are now ready to prove Theorems~\ref{theorem-proposed-1-generic} and~\ref{theorem-proposed-2-generic}.
%

\begin{IEEEproof}[Proof of Theorem~\ref{theorem-proposed-1-generic}]
The proof of the mean square sense convergence claim
follows from Lemma~\ref{theorem-proposed-alg} by applying~\eqref{eqn-lemma-for-proof-of-thm-2}. Namely,
 setting $a:=1-\alpha \mu$ and $v_t:=1-p_{t}^2$, the desired result follows.

 We now prove the almost sure convergence claim.
  By Lemma~\ref{theorem-proposed-alg},
  using $p_k = 1-u_k$, we have:
  \begin{eqnarray*}
  \mathbb E \left[ \|x^{(k)}-x^\bullet\|^2 \right]
 &\leq& 8\,N\, (1-\alpha \mu )^{2 k} D^2 \\
 &+&
 \frac{2\alpha\,{\mathcal{C}_{e}}}{\mu} \sum_{t=1}^k (1-\alpha \,\mu)^{k-t}u_{t-1}.
  \end{eqnarray*}
Now, from~\eqref{eqn-lemma-for-proof-of-thm-2-new},
there exists a positive constant $\mathcal C_u^\prime$ such that,
for all $k=1,2,...$:
\[
\sum_{t=1}^k (1-\alpha\,\mu)^{k-t}u_{t-1} \leq \frac{\mathcal C_u^\prime}{k^{1+\zeta}},
\]
and hence:
 \begin{eqnarray*}
  \mathbb E \left[ \|x^{(k)}-x^\bullet\|^2 \right]
 &\leq& 8\,N\, (1-\alpha \mu )^{2 k} D^2 \\
 &+&
 \frac{2\alpha\,{\mathcal{C}_{e}}\,\mathcal C_u^\prime}{\mu} \frac{1}{k^{1+\zeta}}.
  \end{eqnarray*}
Summing the above inequality over $k=1,2,...$, we obtain that:
\begin{equation}
\label{eqn-for-cheby}
\sum_{k=1}^\infty \mathbb E \left[ \|x^{(k)}-x^\bullet\|^2 \right] < \infty.
\end{equation}
Applying the Chebyshev's inequality and \eqref{eqn-for-cheby},
we conclude that:
\[
\sum_{k=1}^\infty \mathbb P \left( \|x^{(k)}-x^\bullet\| > \epsilon \right) < \infty,
\]
for any $\epsilon>0$. Therefore,
by the first Borel-Cantelli lemma,
$\mathbb P \left( \|x^{(k)}-x^\bullet\| > \epsilon,\,\,\mathrm{infinitely\,often} \right)=0$,
 which finally implies that $x^{(k)}$ converges to $x^\bullet$, almost surely.

\end{IEEEproof}


%
\begin{IEEEproof}[Proof of Theorem~\ref{theorem-proposed-2-generic}]
Consider~\eqref{eqn-to-take-expectation}. Taking expectation:
{\allowdisplaybreaks{
\begin{eqnarray}
\label{eqn-to-take-expectation-2}
\mathbb E \left[ \|x^{(k)}-x^\bullet\| \right]
&\leq& \sqrt{N}(1-\alpha \,\mu)^k 2 D  \\
&+& \alpha
\sum_{t=1}^k (1-\alpha \mu)^{k-t} \sqrt{{\mathcal{C}_{e}} (1-p_{t-1}^2)} \nonumber \\
\label{eqn-to-take-expectation-3}
&\leq& \sqrt{N}(1-\alpha \,\mu)^k 2 D  \\
&+& \alpha
\sum_{t=1}^k (1-\alpha \mu)^{k-t} \sqrt{{\mathcal{C}_{e}}} \sqrt{2} (\sqrt{\delta})^{t}. \nonumber
\end{eqnarray}}}
The first inequality uses $\mathbb E[|u|] \leq (\mathbb E[|u|^2])^{1/2}$.
The second inequality
uses $1-p_{t-1}^2  = (1-(1-\delta^{t}))^2 \leq 2 \delta^{t}$.
 Consider the sum in~\eqref{eqn-to-take-expectation-3}.
  For each $t$, each summand is upper bounded
  by $\eta^k \sqrt{2 {\mathcal{C}_{e}}}$, and so the sum is
  $O(k \eta^k)$. The term $\sqrt{N} (1-\alpha \,\mu)^k 2 D = O(\eta^k)$.
  Hence, the overall right-hand-side in
  \eqref{eqn-to-take-expectation-3} is $O(k \eta^k) = O((\eta+\epsilon)^k)$, which completes the proof.
\end{IEEEproof}
\begin{remark}
\label{remark-hidden-constant}
The proof of Theorem~\ref{theorem-proposed-2-generic}
also determines the constant in the convergence rate. From the
above proof, substituting the expression for
${\mathcal{C}_{e}}$ in~\eqref{eqn-C-e-value}, it is straightforward to observe that, for all $k=1,2,...$:
\begin{equation*}
\mathbb E \left[ \|x^{(k)}-x^\bullet\| \right] \leq 12\,
\max \left\{ \sqrt{N}\,D,\,\,\frac{\alpha\,\sqrt{N}\,G}{p_{\mathrm{min}}(1-\beta)}\right\}\,k\,\eta^k.
\end{equation*}
\end{remark}

\section{Simulations}
\label{section-simulations}
We provide simulations on the problem of learning a linear classifier via logistic loss,
both on synthetic and real data sets.
Simulations demonstrate that our proposed idling strategy significantly
reduces the total cost (both communication and computational costs), when compared with standard distributed gradient
 method where all nodes work at all iterations.
 Simulations also demonstrate the method's high degree of robustness to
 asynchrony and its benefits over gossip-based strategies for solving~\eqref{eqn-opt-prob-original}.

We consider distributed learning of
 a linear classifier via logistic loss, e.g.,~\cite{BoydADMoM}.
 Each node $i$ possesses $J$ data samples
 $\{a_{ij}, b_{ij}\}_{j=1}^{J}$. Here, $a_{ij} \in {\mathbb R}^{d-1}$, $d \geq 2$, is a feature vector, and
$b_{ij} \in \{-1,+1\}$ is its class label.
 We want to learn a vector $x=(x_1^\top,x_0)^\top$,
$x_1 \in {\mathbb R}^{d-1}$, and
$x_0 \in {\mathbb R}$, $d \geq 1$, such that
the corresponding linear classifier $\mathrm{sign} \left(H_x(a)\right) = \mathrm{sign}
\left( x_1^{\top}a+x_0\right)$ minimizes the total surrogate loss with $l_2$ regularization:
\begin{equation}
\label{eqn-sim-prob}
\sum_{i=1}^N \left(\sum_{j=1}^J \mathcal{J}_{\mathrm{logis}} \left(  b_{ij} H_x(a_{ij})\right) + \frac{1}{2}\mathcal{R} \|x\|^2\right),
\end{equation}
subject to a prior knowledge that $\|x\| \leq \mathcal{M}$, where $\mathcal M>0$ is a constant.
 Here, $\mathcal{J}_{\mathrm{logis}}(\cdot)$ is the logistic loss $\mathcal{J}_{\mathrm{logis}}(\alpha) = \log (1+e^{-\alpha})$,
and $\mathcal{R} $ is a positive regularization parameter. Clearly, problem~\eqref{eqn-sim-prob} fits the generic framework in~\eqref{eqn-opt-prob-original} with
 $f_i(x)=\sum_{j=1}^J \mathcal{J}_{\mathrm{logis}} \left(  b_{ij} H_x(a_{ij})\right) + {\mathcal{R}} \|x\|^2/2$,
 $f(x)=\sum_{i=1}^N f_i(x)$, and $\mathcal{X} = \{x \in {\mathbb R}^4:\,\|x\| \leq \mathcal M\}$.
   A strong convexity constant of the $f_i$'s $\mu$ can be taken
 as $\mu = {\mathcal{R}}$, while a Lipschitz
 constant $L$ can be taken as $\frac{1}{4 N} \|\sum_{i=1}^N \sum_{j=1}^J c_{ij}\,c_{ij}^\top\| + {\mathcal R}$,
 where $c_{ij} = (b_{ij}\,a_{ij}^\top, b_{ij})^\top$.

With all experiments, we test the algorithms on a connected network with $N=50$ nodes and $214$ links,
 generated as a random geometric graph: we place nodes
 randomly (uniformly) on a unit square, and the node pairs whose distance is less
than a radius are connected by an edge.

All our simulations are implemented in Matlab
on a single standard personal computer. Hence, the network
itself is also emulated over a single computer.
An interesting future step is to
 run the proposed algorithm over a distributed environment (e.g., a cluster with 50-100 cores)
 where actual operation imperfections may reveal further
 real implementation effects.

\textbf{Experiments on synthetic data}. In the first set of experiments, we generate data and set the algorithm parameters as follows.
 Each node $i$ has $J=2$ data points whose dimension is $d-1=3.$
We generate the $a_{ij}$'s independently over $i$ and $j$; each entry of $a_{ij}$ is drawn independently
from the standard normal distribution.
We generate the ``true'' vector $x^\star=((x_1^\star)^\top, x_0^\star)^\top$
by drawing its entries independently from standard normal distribution.
Then, the class labels are generated as $b_{ij}=\mathrm{sign} \left( (x^\star_1)^\top a_{ij}+x^\star_0+\epsilon_{ij}\right)$, where
$\epsilon_{ij}$'s are drawn independently from normal distribution with zero mean and
standard deviation~$0.1$. The obtained corresponding strong convexity parameter $\mu=0.1$,
and the Lipschitz constant~$L \approx 0.69$. Further, we set $\mathcal M = 100$
 and $\mathcal{R} =0.1$.

With both algorithms, we initialize $x_i(0)$ to $\mathcal{P}_{\mathcal{X}}(h_i)$,
where the $h_i$'s, $i=1,...,N$, are generated mutually independently, and the entries of each $h_i$ are generated mutually independently from the uniform distribution on $[-50,+50]$. We utilize the Metropolis weights, e.g.,~\cite{BoydFusion}. With the proposed method, we set $p_k = 1-\delta^{k+1}$,
$k=0,1,...,$ and $\delta =(1-\alpha\,\mu)^2$.

As an error metric, we use
   the relative error in the objective function averaged across nodes:
    \[
   \frac{1}{N} \sum_{i=1}^N \frac{f(x_i^{(k)})-f^\star}{f^\star}, \,\,f^\star>0,\]
where $f^\star$ is evaluated numerically via the (centralized) projected gradient method.
 With the proposed method, we run $100$ simulations and consider both the
 average (empirical mean) relative error (averaged across $100$ simulation runs with
different instantiations of node activations, i.e., variables $z(k)$) and the relative error's histograms.

We compare the two methods with respect to the total cost
(which equals the total number of activations across all nodes), where a unit cost
 corresponds to a single node activation at one iteration; we also include the
 comparisons with respect to
the number of iterations~$k$.
We consider two different values of step-sizes, $\alpha \in \{\frac{1}{250\,L},\,
\frac{1}{50\,L}\}$, which correspond to different achievable accuracies by both methods.

\begin{figure}[htb]
\begin{minipage}[b]{1.0\linewidth}
  \centering
  \includegraphics[height=2.2 in,width=2.7 in]{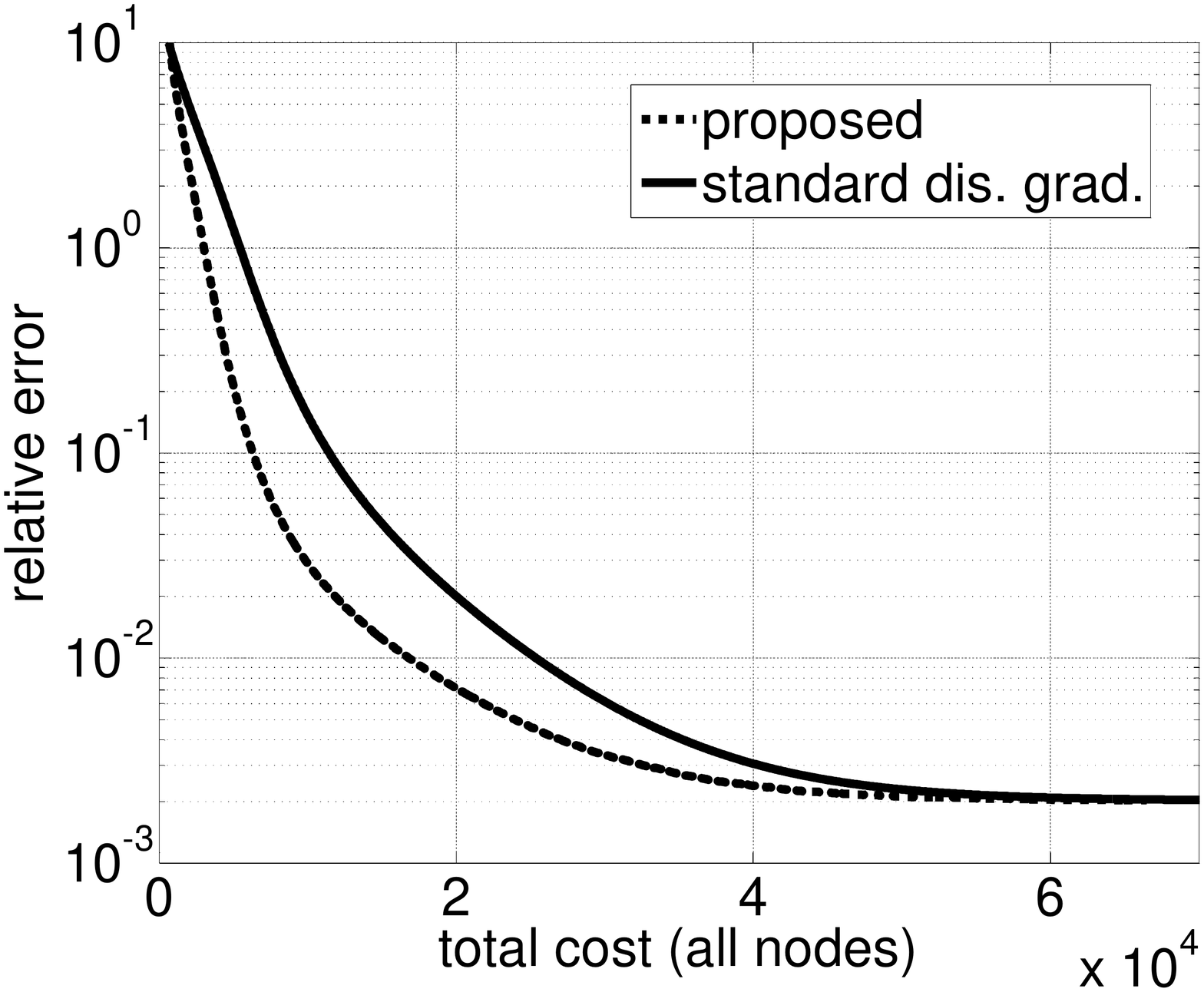}
  \centerline{(a) Average relative error vs. total cost (all nodes).}\medskip
\end{minipage}
\begin{minipage}[b]{\linewidth}
  \centering
 \includegraphics[height=2. in,width=2.7 in]{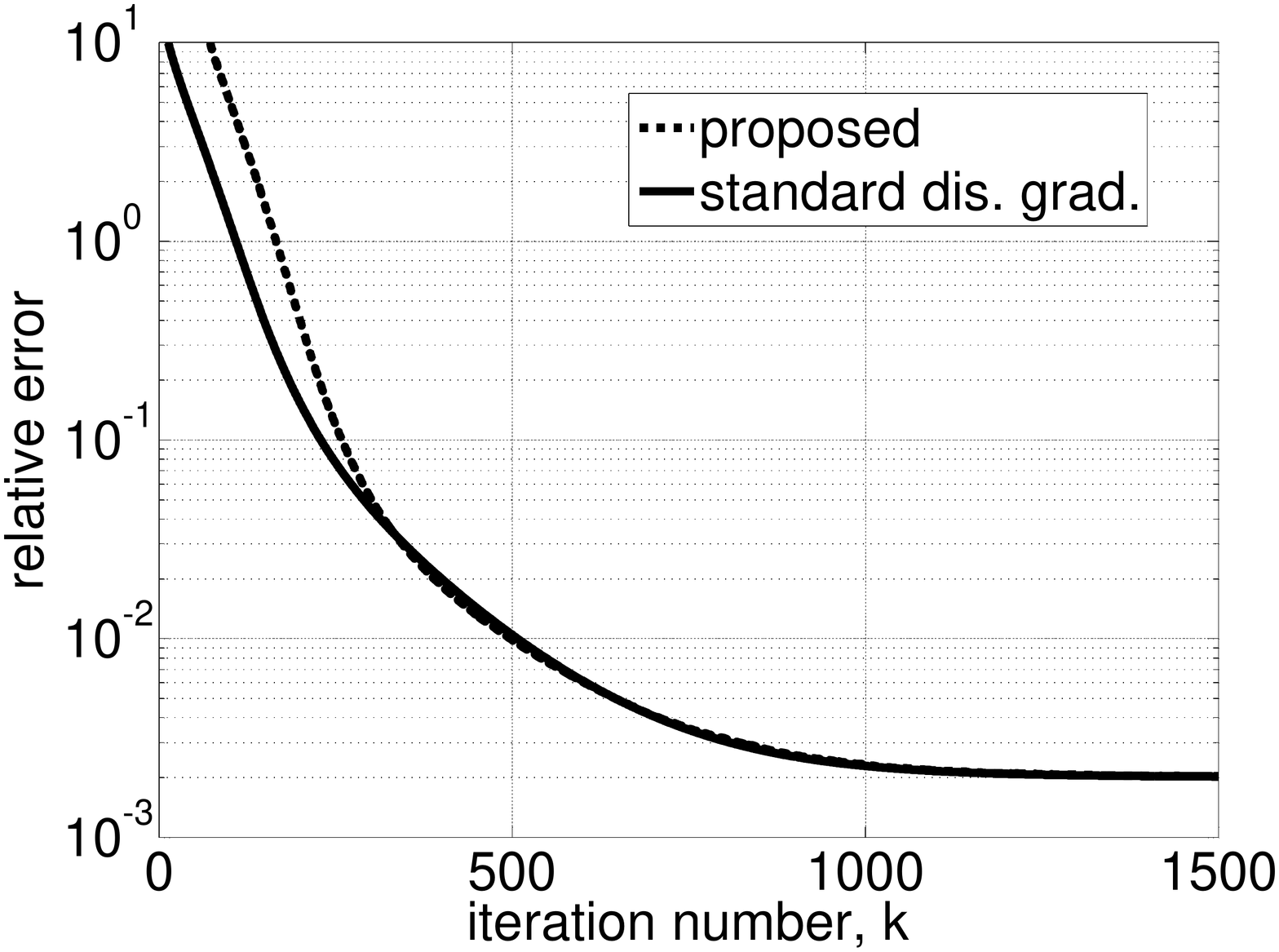}
  \centerline{(b) Average relative error vs. number of iterations.}\medskip
\end{minipage}
\begin{minipage}[b]{1.0\linewidth}
  \centering
  \includegraphics[height=1.9 in,width=2.7in]{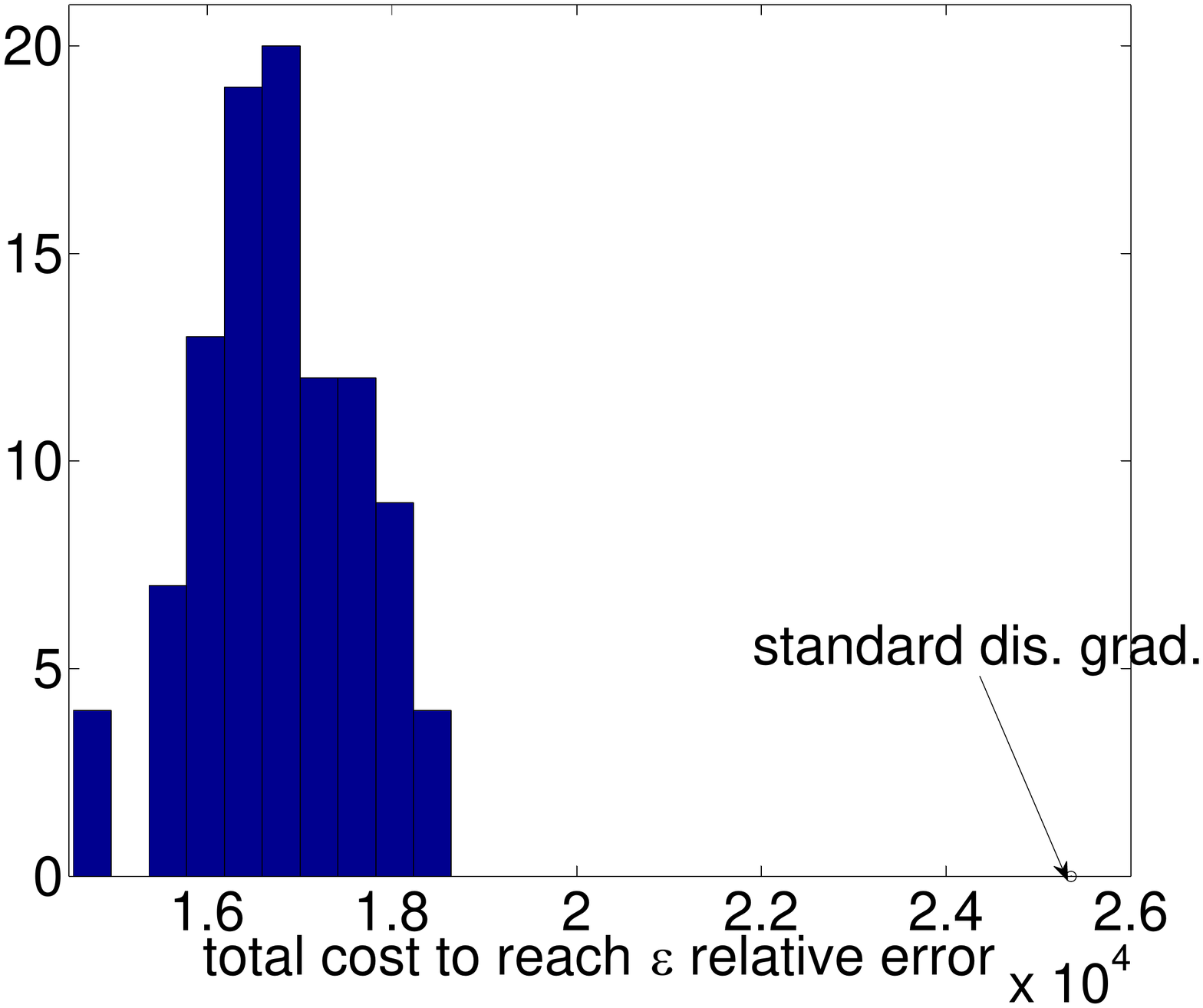}
  \centerline{(c) Histogram: total cost to reach rel. err. $0.01.$}\medskip
\end{minipage}
\begin{minipage}[b]{1.0\linewidth}
  \centering
  \includegraphics[height=2.0 in,width=2.5 in]{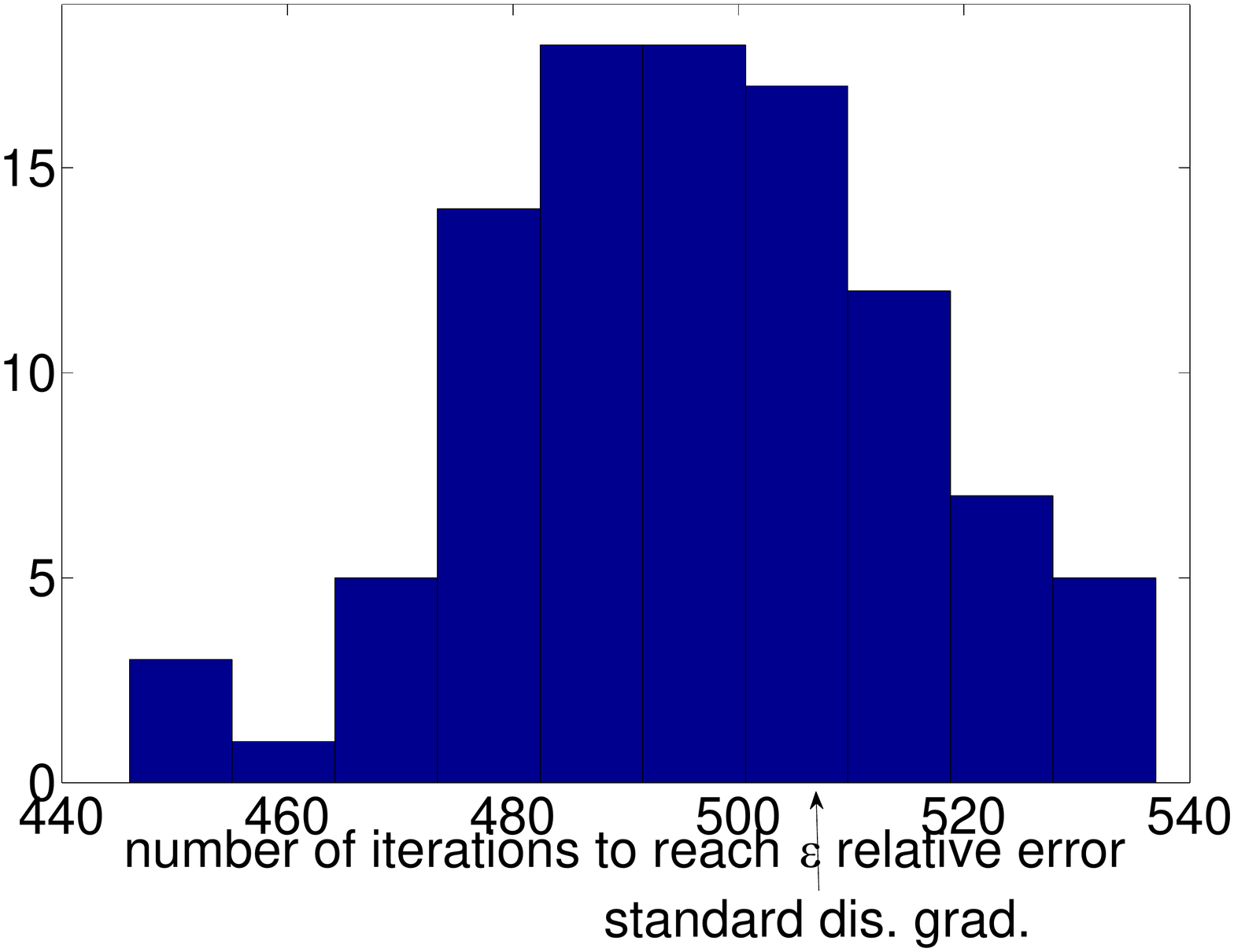}
\centerline{(d) Histogram: $\#$iterations to reach rel. err. $0.01.$}\medskip
\end{minipage}
\caption{Comparison of the proposed and standard distributed gradient methods for the synthetic
data set and $\alpha=
\frac{1}{50\,L}$. In Figures~(c) and~(d), the arrows indicate the performance
of the standard distributed gradient method: total cost~$\approx25,500$ (Figure~(c));
and number of iterations~$\approx507$ (Figure~(d)).}
\label{Figure1}
\end{figure}

\begin{figure}[htb]
\begin{minipage}[b]{1.0\linewidth}
  \centering
  \includegraphics[height=2.2 in,width=2.7 in]{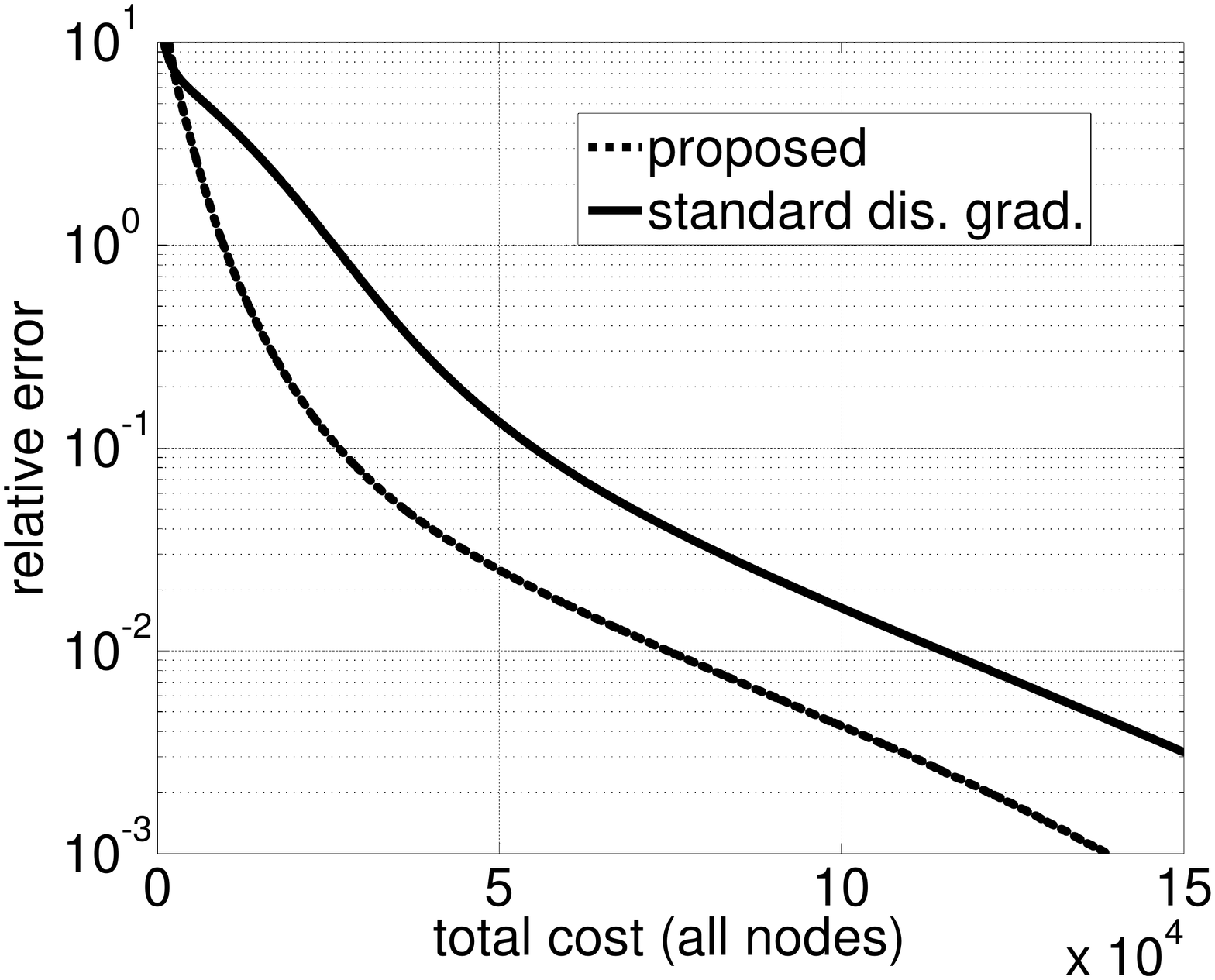}
  \centerline{(a) Average relative error vs. total cost (all nodes).}\medskip
\end{minipage}
\begin{minipage}[b]{1.0\linewidth}
  \centering
  \includegraphics[height=2.2 in,width=2.7 in]{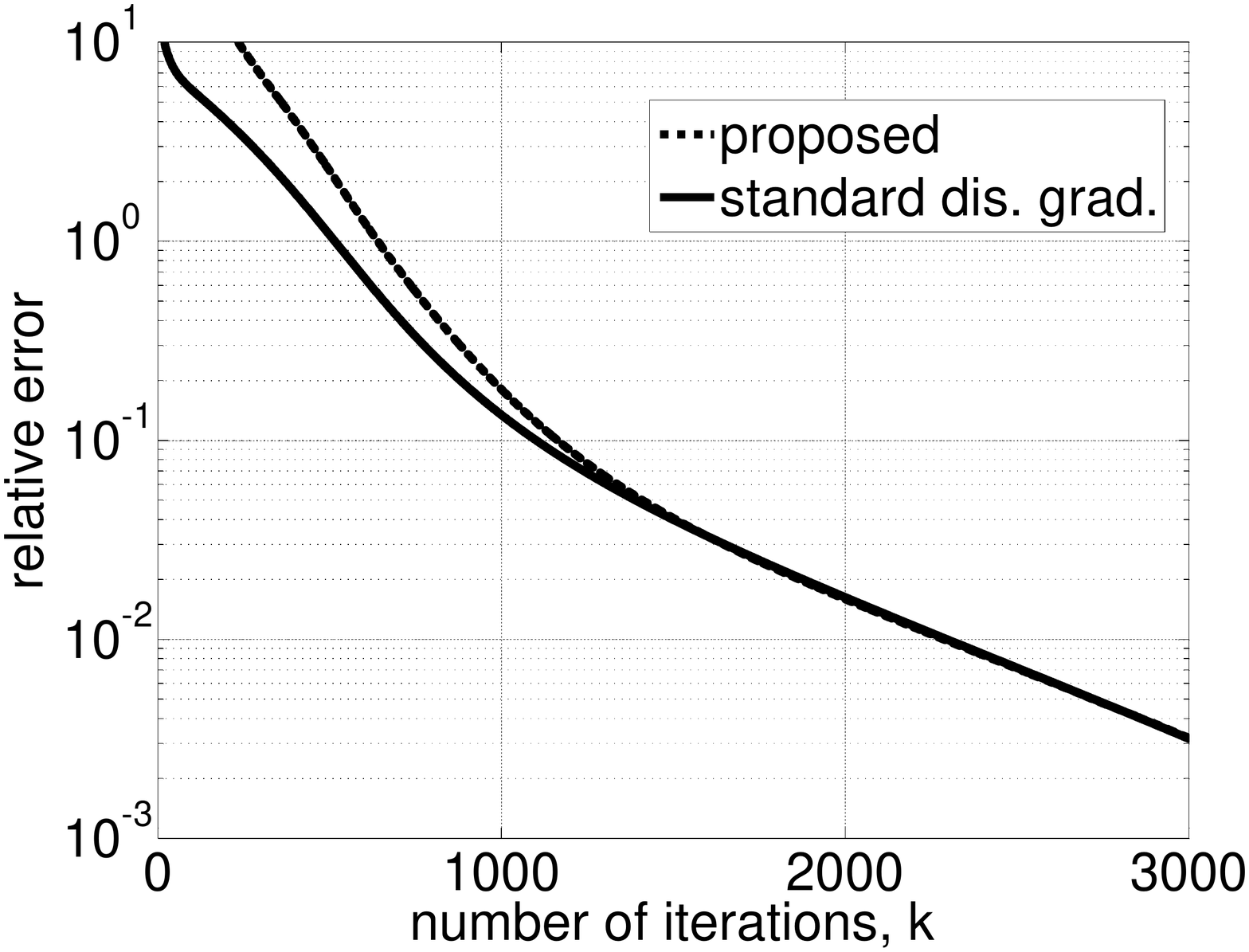}
  \centerline{(b) Average relative error vs. number of iterations.}\medskip
\end{minipage}
\begin{minipage}[b]{1.0\linewidth}
  \centering
  \includegraphics[height=1.8 in,width=2.4 in]{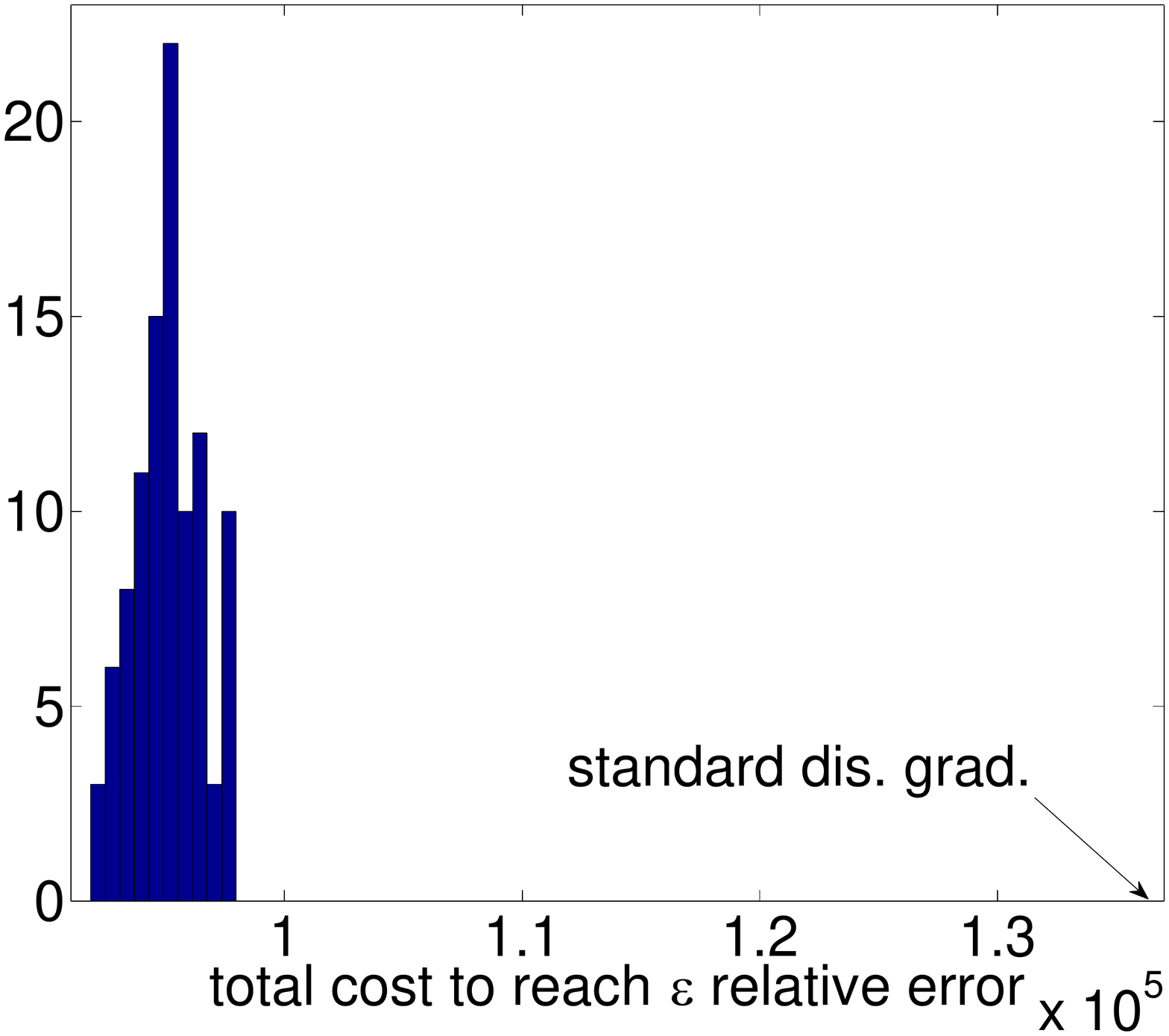}
  \centerline{(c) Histogram: total cost to reach rel. err. $0.005.$}\medskip
\end{minipage}
\begin{minipage}[b]{1.0\linewidth}
  \centering
  \includegraphics[height=1.8 in,width=2.5 in]{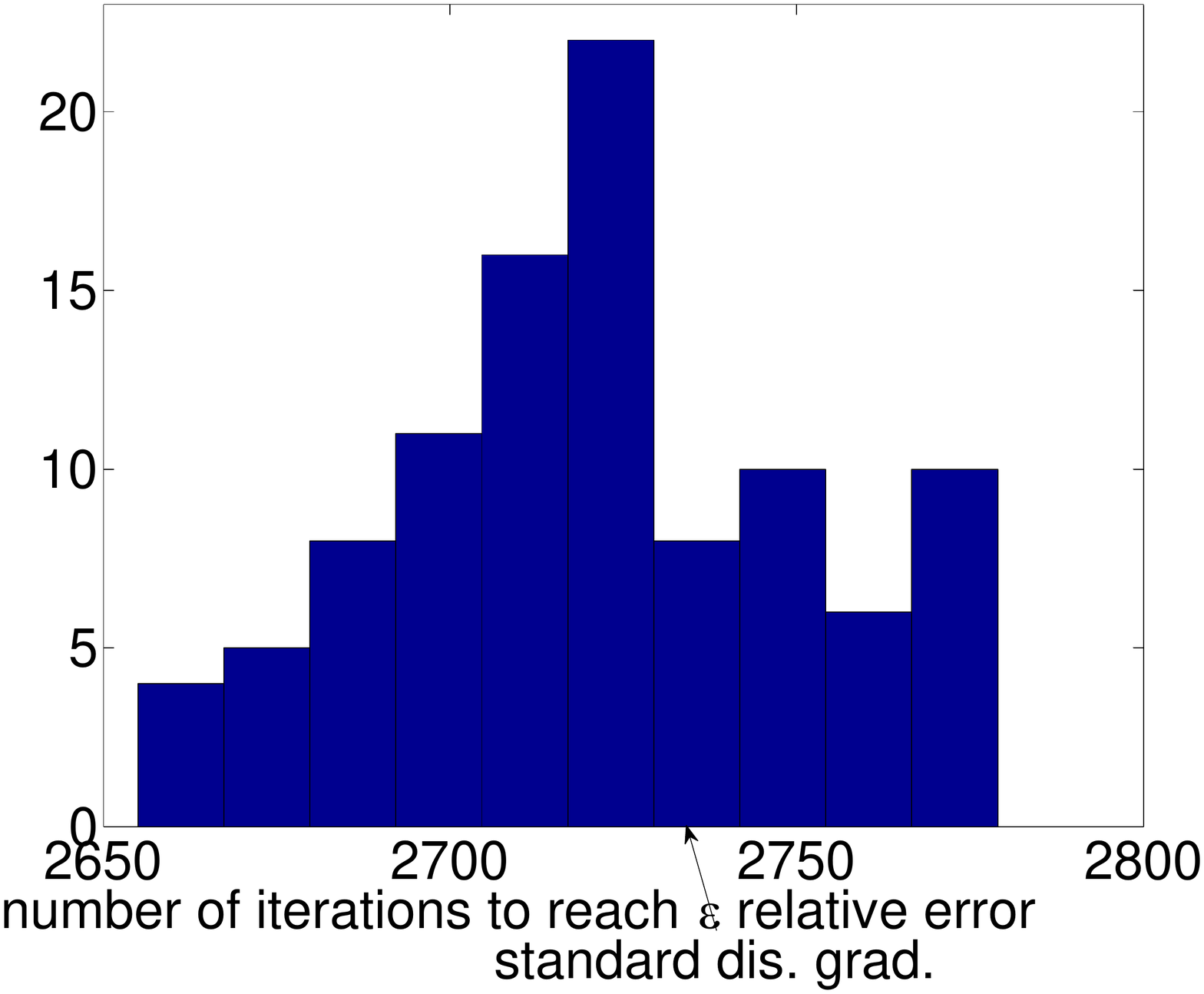}
  \centerline{(d) Histogram: total cost to reach rel. err. $0.005.$}\medskip
\end{minipage}
\caption{Comparison of the proposed and standard distributed gradient methods for $\alpha=
\frac{1}{250\,L}$. In Figures~(c) and~(d), the arrows indicate the performance
of the standard distributed gradient method: total cost~$\approx137,000$ (Figure~(c));
and number of iterations~$\approx2,740$ (Figure~(d)).}
\label{Figure2}
\end{figure}

Figure~\ref{Figure1} compares the proposed and standard distributed gradient methods
for~$\alpha = 1/(50L)$. We can see that the proposed method significantly reduces total cost
 to reach a certain accuracy,
 while at the same time it does not induce an overhead in the total number of iterations.
 For example, consider Figure~\ref{Figure1}~(a)
 which plots the relative error averaged over $100$ simulation runs
 versus total cost. We can see that, to achieve relative error~$\epsilon = 0.01$,
 the proposed method has on average the total cost around $16,700$, while the standard distributed gradient method requires around $25,500$,
 which gives the relative savings of about~$33\%$. At the same time,
 the two methods require practically the same number of iterations (Figure~\ref{Figure1}~(b)).
 Figure~\ref{Figure1}~(c) shows the
 histogram for the proposed method of the total
 cost to achieve relative error $\epsilon=0.01$,
 where an arrow indicates the cost of the standard distributed gradient method
  (which equals $25,500$). Figure~\ref{Figure1}~(d) repeats the study for the total number of iterations.
     Figure~\ref{Figure2} shows the
   comparisons for~$\alpha = 1/(250 L)$, and
   it shows histograms to reach~$\epsilon = 0.005$, again demonstrating
   very significant improvements.
   To reach relative error~$\epsilon = 0.005$,
   the proposed method requires the total cost around~$90,000$,
   while the standard distributed gradient method takes~$137,000$,
   which corresponds to the proposed method's savings of about~$44\%$.

\begin{figure}[htb]
\begin{minipage}[b]{1.0\linewidth}
  \centering
  \includegraphics[height=2.0 in,width=2.7 in]{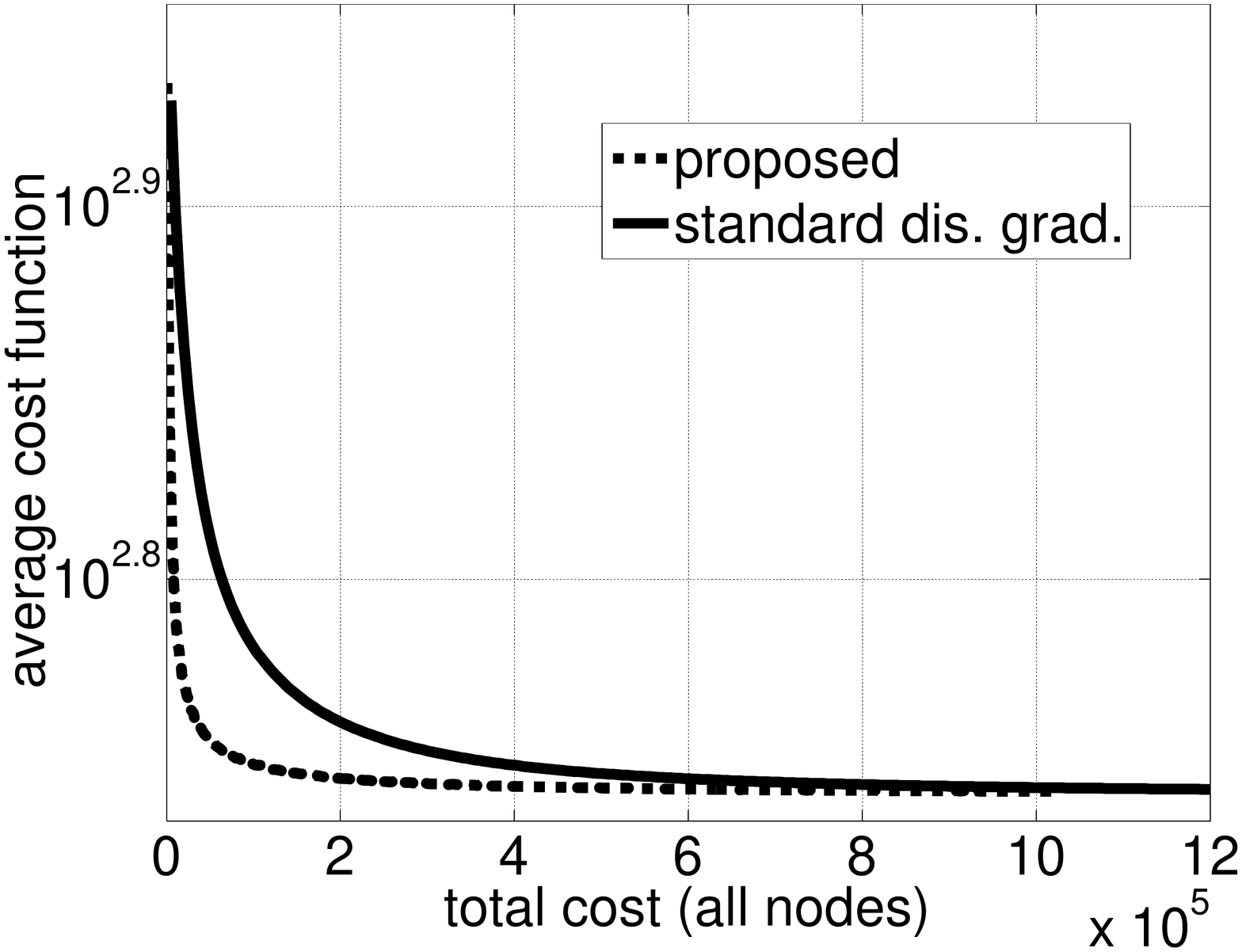}
  \centerline{(a) Average cost function vs. total cost for ``a1a.''}\medskip
\end{minipage}
\begin{minipage}[b]{1.0\linewidth}
  \centering
  \includegraphics[height=2.0 in,width=2.7 in]{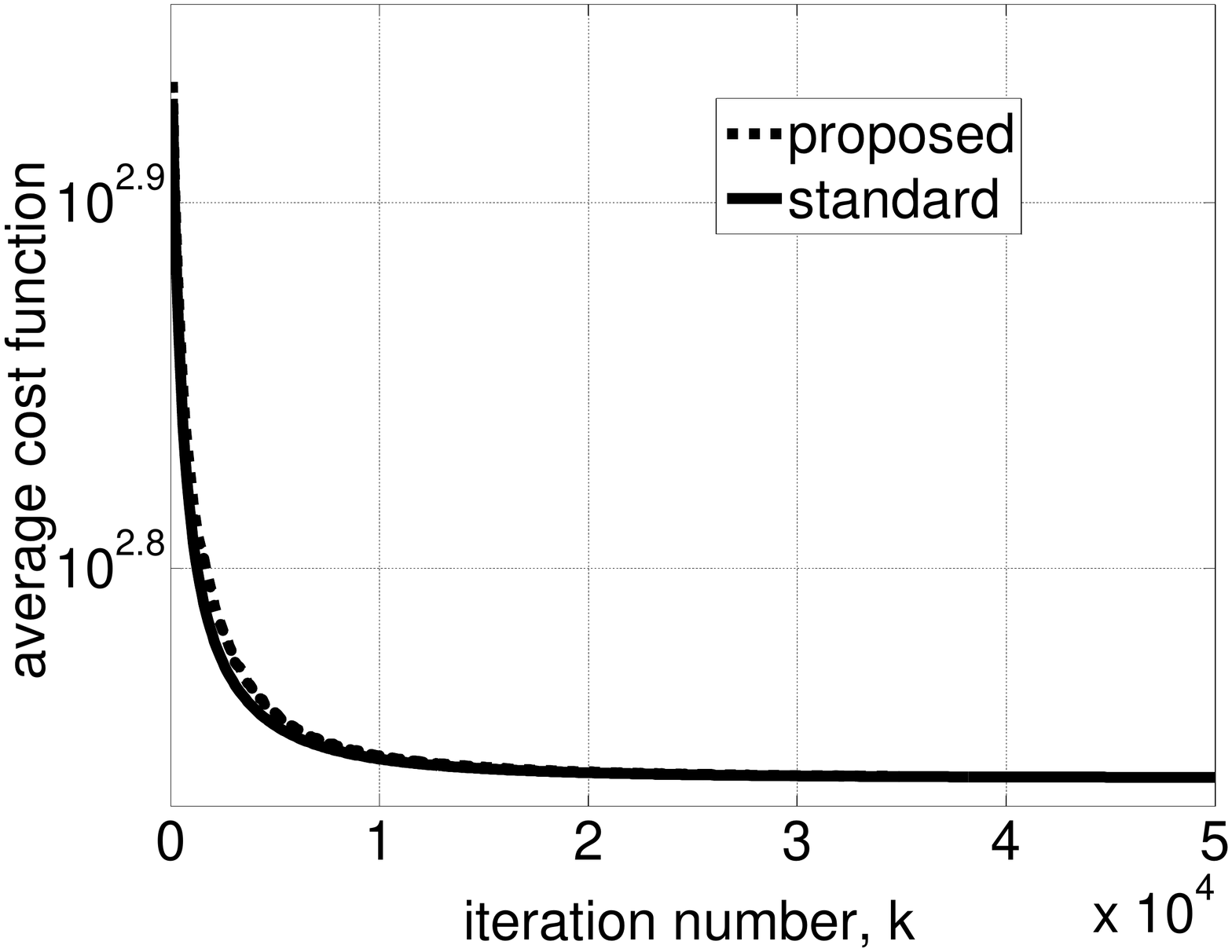}
  \centerline{(b) Average cost function vs. number of iterations for ``a1a.''}\medskip
\end{minipage}
\begin{minipage}[b]{1.0\linewidth}
  \centering
  \includegraphics[height=2. in,width=2.7 in]{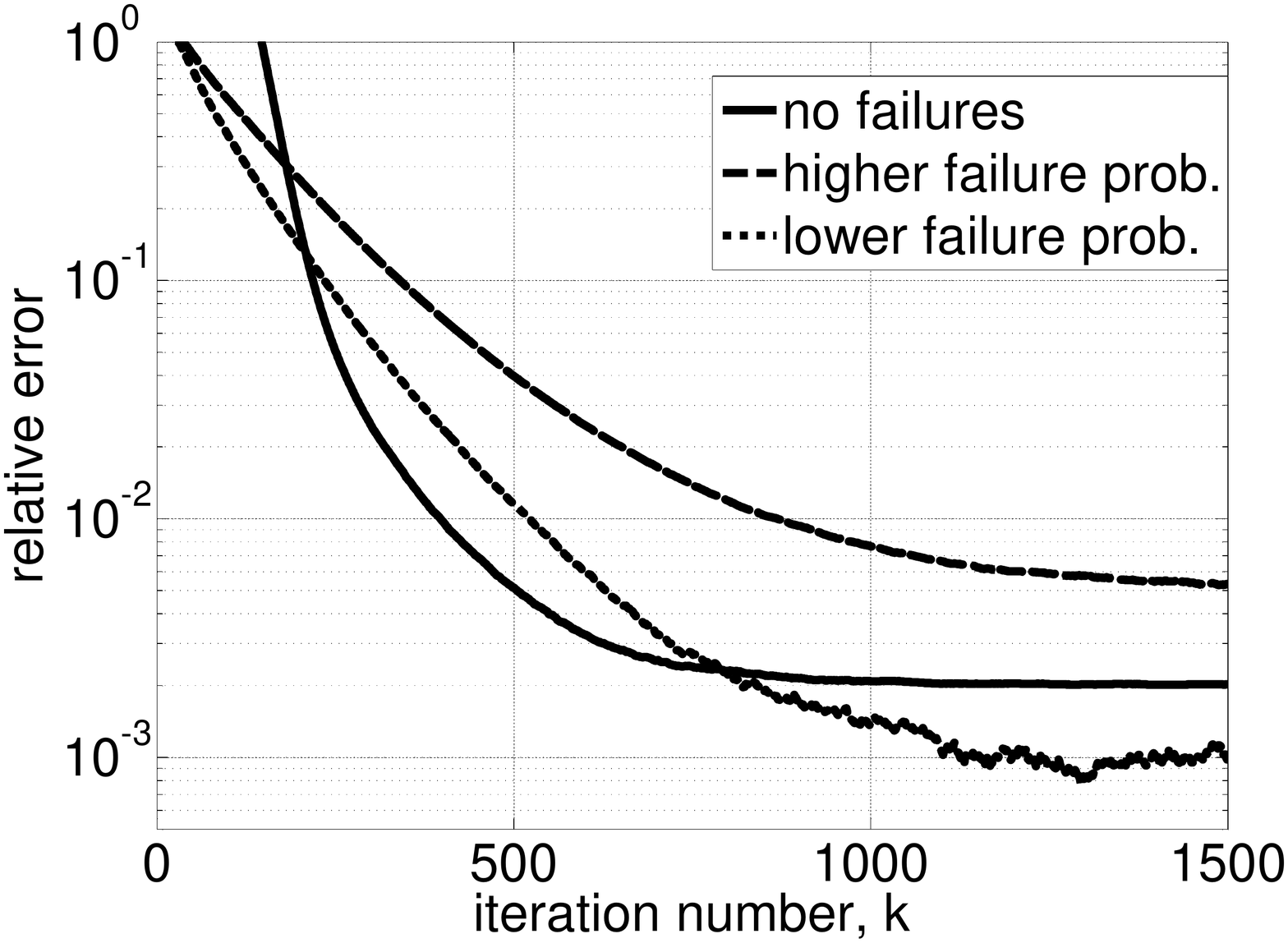}
  \centerline{(c) Average relative error vs. number of iterations.}\medskip
\end{minipage}
\begin{minipage}[b]{1.0\linewidth}
  \centering
  \includegraphics[height=1.9 in,width=2.7 in]{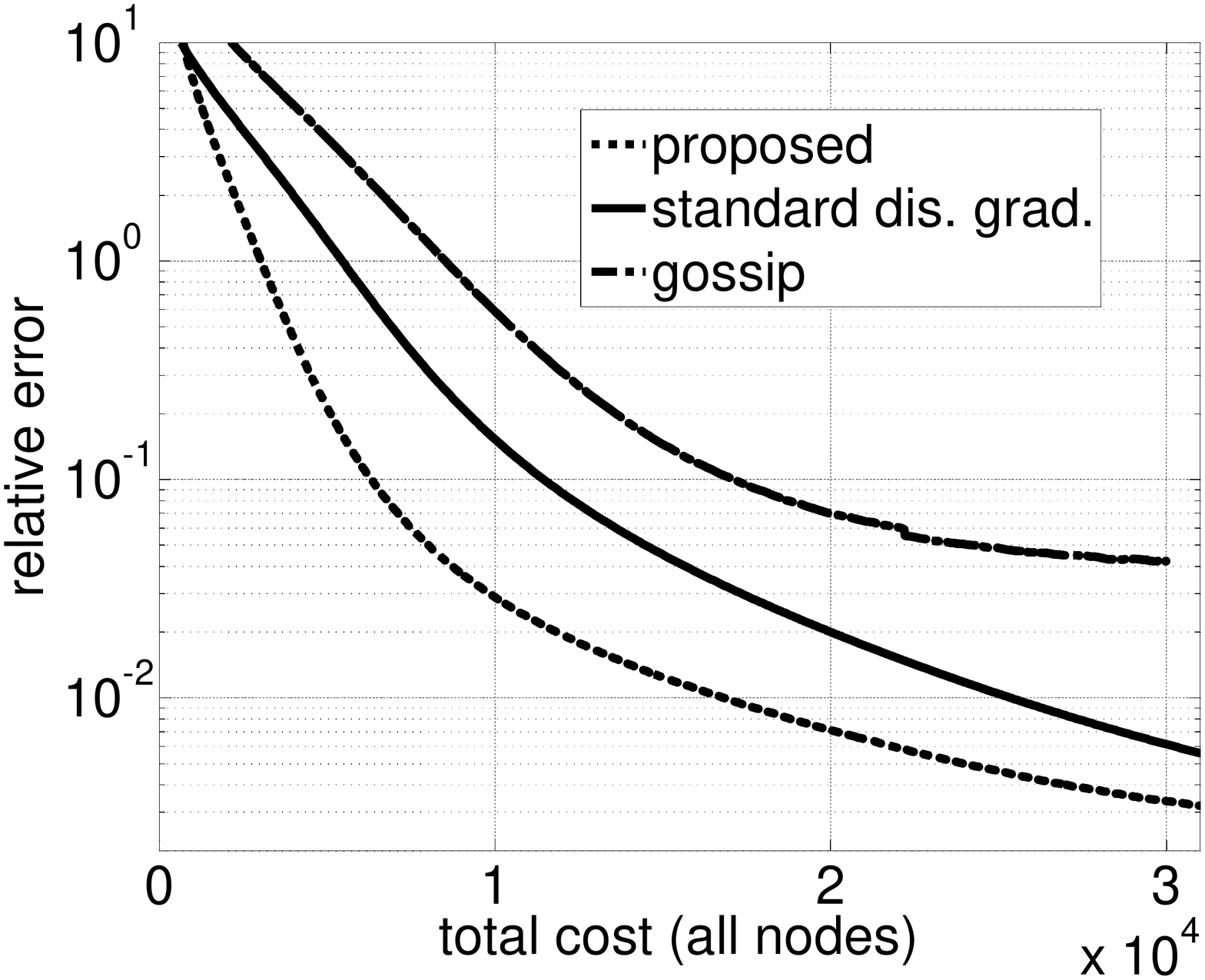}
  \centerline{(d) Average relative error vs. total cost.}\medskip
\end{minipage}
\caption{Figures~(a) and~(b):
Comparison of the proposed and standard distributed gradient methods for data set
 ``a1a;'' Figure~(c): Effect of failures on the proposed method for
 the synthetic data set;
 and Figure~(d): Comparison of
 the proposed method with
 the gossip-based scheme in~\cite{nedic-gossip}
  for the synthetic data set.
 }
\label{Figure3}
\end{figure}

%

\textbf{Experiments on a real world data set}. In the second set of experiments, we consider the same network with $50$ nodes
and test the algorithms on data set ``a1a'' which
we downloaded from the repository: http://www.csie.ntu.edu.tw/~cjlin/libsvm/.
   There are $N\,J=1,600$ data points\footnote{There are actually $1,605$
 points but we discarded $5$ of them to evenly distributed data points across nodes.}
 ($J = 32$ per node) of dimension $d-1=119$ (optimization variable dimension is~$120$).
 We set~$\mathcal M = 100$
 and $\mathcal{R} =0.1$, $\mu={\mathcal R}$, and $L=
 \frac{1}{4} \max_{i=1,...,N} \|\sum_{j=1}^J c_{ij}\,c_{ij}^\top\| + {\mathcal R}.$

We use all the system and algorithmic parameters the same as in the first set of experiments,
except the following. With both methods,
we initialize $x_i^{(0)}$ to zero, for all~$i$. With the proposed method, we set $p_k = \max\{1-\delta^{k+1},\,0.1\}$,
$k=0,1,...,$ and $\delta =\min\{(1-\alpha\,\mu)^2,\,0.99999\}$.
(See the discussion in the last paragraph of Section~{III}.)
As the error metric, we use
   the average cost function (averaged across nodes)
     \[
   \frac{1}{N}\sum_{i=1}^N f\left( x_i^{(k)}\right).
    \]
   %
 With the proposed method,
 we run one sample path realization.

Figure~\ref{Figure3}, a) and b), compares the proposed and standard distributed gradient methods for the ``a1a''
data set and step size~$\alpha = 1/(50L)$. We can that the proposed method reduces the total
 cost at least $3$ times, while incurring a very low
 overhead in the total number of iterations.

\textbf{Modeling and testing asynchronous operation}. In applications like, e.g., WSNs,
 accounting for asynchrony in the algorithm's operation is highly relevant.
 In such scenarios, when node~$i$ decides to
 activate at iteration $k$ and transmit a message to node $j$, this message may be lost, due to, e.g.,
 packet dropouts in WSNs. In addition, an active node
 may fail to calculate the local gradient at iteration~$k$, because
 the actual calculation may take longer than the time slot allocated to iteration~$k$,
  or due to unavailability of sufficient computational resources at~$k$. Therefore,
  under asynchrony, the schedule of the realized inter-neighbor communications and
  local gradient evaluations is not under the full control of networked nodes.

We introduce the following model. At each link $\{i,j\} \in E$
 and each~$k=0,1,...,$ let $\widehat{z}_{\{i,j\}}^{(k)}$ be a binary
   random variable which takes value one if the communication link
   is online and zero otherwise; let $\widehat{p}_{ij}:
   =\mathbb P \left( \widehat{z}_{\{i,j\}}^{(k)}=1\right)$.
   Therefore, variable $\widehat{z}_{\{i,j\}}^{(k)}$ models
   a failure of link $\{i,j\}$ at~$k$.
   Similarly, for each node~$i$, introduce a binary random variable
   $\widehat{z}_i^{(k)}$, which takes the value one
   if the calculation of $\nabla f_i \left( x_i^{(k)}\right)$
    is successful and zero otherwise.
    We let ${\widehat p}_i: = \mathbb P \left( \widehat{z}_{i}^{(k)}=1\right)$.
     Variable~$\widehat{z}_{i}^{(k)}$ hence models
   failure of node $i$'s gradient calculation at~$k$.
      (As before, each node $i$ activates
     if $z_i^{(k)}=1$ and stays idle if $z_i^{(k)}=0.$)
     We assume that the variables $\widehat{z}_{\{i,j\}}^{(k)}$ are independent
     both across links and across iterations; likewise, the
     $\widehat{z}_{i}^{(k)}$'s are independent
     both across nodes and across iterations; and that
    the node activations, the link failures, and the gradient calculation
     failures are mutually independent processes. Note that
     $z_i^{(k)}$ is in control of node~$i$, while the $\widehat{z}_{\{i,j\}}^{(k)}$'s
      and $\widehat{z}_{i}^{(k)}$'s are governed ``by nature.''
   The update of node $i$ and iteration~$k$ is as follows.
   If~$z_i^{(k)}=0$, node $i$ stays idle;
   else, if~$z_i^{(k)}=1$, we have:
  \begin{eqnarray}
  \label{eqn-update-one-node-asynchronous}
  x_i^{(k+1)} &=& \mathcal{P}_{\mathcal X} \{ \,( 1-\sum_{j \in \Omega_i}
  z_j^{(k)} \widehat{z}_{\{i,j\}}^{(k)}C_{ij} \,) x_i^{(k)} \\
  &+& \sum_{j \in \Omega_i} C_{ij} \,z_j^{(k)} \widehat{z}_{\{i,j\}}^{(k)}\, x_j^{(k)} \nonumber \\
  &-& \frac{\alpha}{p_k} z_i^{(k)} \widehat{z}_{i}^{(k)} \nabla f_i(x_i^{(k)})\,\}. \nonumber
  \end{eqnarray}
Note that we assume that nodes do not have prior knowledge on the
asynchrony parameters~$\widehat{p}_i$'s and~$\widehat{p}_{\{i,j\}}$'s.

We provide a simulation example on the synthetic data set and the~$50$-node network
considered before with the same simulation parameters and~$\alpha=1/(50 L)$.
 Each $\widehat{p}_{\{i,j\}}$ is set to $0.5$, while
 for the $\widehat{p}_i$'s we consider two scenarios:
 1) lower failure probabilities, where one half of the nodes
 has $\widehat{p}_i=0.9$ and the other half has $\widehat{p}_i=0.5$;
  and 2) higher failure probabilities, where a half of the nodes
 has $\widehat{p}_i=0.9$ and the other half has $\widehat{p}_i=0.1$.
  Note that the latter scenario
 corresponds to rather severe conditions, as one half of the nodes
 successfully computes the gradients only with~$0.1$ probability.

Figure~\ref{Figure3}~(c) shows the performance of
the proposed method for three scenarios: no failures,
lower failure probabilities, and higher failure probabilities.
 It shows the empirical mean
of the relative error averaged across $100$
 iterations (higher
 failure probabilities are
 shown with a dashed line, and
 lower failure probabilities with a dotted line.)
 We can see that the proposed algorithm exhibits
 a very strong resilience to asynchrony.
  First consider the higher failure probabilities scenario
   (dashed line in Figure~\ref{Figure3}~(c)).
 We can see that, despite the severe conditions,
 the proposed algorithm still converges close to
 the solution, naturally
 with a decreased convergence rate and with
  a moderately increased limiting error.
  Now, consider the
  lower failure probabilities scenario
   (dotted line in Figure~\ref{Figure3}~(c)).
   The proposed algorithm again generally
   slows down convergence, as it is expected.
   However, interestingly, it actually achieves a higher
   degree of accuracy asymptotically than
   under the synchronous scenario.
   This is explained as follows.
   The effective step-size of node $i$
    with algorithm~\eqref{eqn-update-one-node-asynchronous}
    equals~$\frac{\alpha}{p_k} z_i^{(k)} \widehat{z}_{i}^{(k)}$,
    which is on average~$\alpha \widehat{p}_i$.
    Hence, in a sense,
   $\widehat{p}_i$ has the effect of
   decreasing the step-size. The
   step-size decrease has the known effect of
   slowing down convergence rate but
   improving the asymptotic accuracy,
   as confirmed in Figure~\ref{Figure3}~(c).
     The improved asymptotic
     accuracy indeed occurs
     as long as the~$\widehat{p}_i$'s
  are not mutually too different. When the
  $\widehat{p}_i$'s are mutually too far apart,
  different nodes effectively use
  very different step-sizes (which equal to $\alpha \widehat{p}_i$),
  and this disbalance makes a negative effect on
  both the convergence speed and on the asymptotic accuracy -- as
  confirmed in Figure~\ref{Figure3}~(c)
  for the higher failure probabilities case.

\textbf{Comparison with a gossip-based scheme}.
 To further corroborate the benefits of the proposed idling scheme
with increasing activation probabilities, we
compare it  -- on the synthetic data set and $\alpha=1/(50 L)$ --
with the gossip-based scheme in~\cite{nedic-gossip}.
Figure~3~(d) plots the relative error, averaged
over $20$ simulation runs, versus iteration number.
(Gossip is shown in a dash-dot line.)
With both methods,
we use the same step-size parameter~$\alpha$.
 We can see that the proposed scheme outperforms gossip.
  Most notably, the gossip scheme has a larger steady state error.
We explain why this happens. Namely, with gossip,
only two nodes (out of $N$) are active at all iterations.
 This means that, essentially,
 the gossip-based scheme behaves as an incremental gradient method (more precisely, a mini-batch)
 gradient method, where the full gradient (which equals the sum
 of $N$ local nodes functions' gradients) is at all times
 approximated with the sum of two local gradients.
 Therefore, the gossip-based scheme incurs
 an increased steady state error,
 for a similar reason as the fact why
 the (centralized) incremental gradient method
 with a constant step size does not converge to the exact solution.
  In contrast, our method essentially behaves as
  a full gradient method, thus leading to a higher accuracy.


\section{Discussion and extensions}
\label{section-extensions}
Subsection~\ref{subsection-relaxing-strong-cvx-diff} extends our results
to the case of convex costs which
do not have to be differentiable nor
strongly convex, while Subsection~{VI-B}
 provides a quantification of
 the gains in the total cost for a special case of quadratic costs
 with identity Hessians. Proofs of the
 results in the current section can be found in the Appendix.

\subsection{Relaxing strong convexity and differentiability}
\label{subsection-relaxing-strong-cvx-diff}
We assumed throughout the previous part of the paper that the $f_i$'s are strongly convex and have Lipschitz continuous gradients. We now extend our results to more generic cost functions, when these two assumptions are relaxed.
 Specifically, we now let each $f_i:\,{\mathbb R}^d \rightarrow \mathbb R$ be convex and Lipschitz over
 set $\mathcal X$, i.e., for all $i$, there holds:
 \begin{equation}
 \label{eqn-lip-f}
 |f_i(x)-f_i(y)| \leq G\,\|x-y\|,\,\,\forall x,y \in \mathcal X,
 \end{equation}
 for a nonnegative constant~$G$. We continue to assume that $\mathcal X$ is convex and compact,
 so \eqref{eqn-lip-f} is satisfied for any convex function, e.g., \cite{duchi}.
   Optimization problem~(1) is solvable under this setting.

The proposed algorithm~\eqref{eqn-update-one-node} generalizes straightforwardly: at node
$i$ and iteration~$k$,
 gradient $\nabla f_i \left( x_i^{(k)} \right)$
  is replaced with an arbitrary subgradient from
   the subdifferential set of $f_i$ at $ x_i^{(k)}$, $\partial f_i \left( x_i^{(k)} \right)$.
   We note that Lemmas~5 and~9 continue to hold here as well.

Before presenting our result on the modified algorithm~\eqref{eqn-update-one-node}, we recall that
the standard distributed gradient method achieves
for the setting assumed here the following performance.
Define, for each node $i$, the running average:
\[
x_{i,\mathrm{ra}}^{(k)} = \frac{1}{k}\sum_{t=0}^{k-1}x_i^{(t)},\,\,k=1,2,...
\]
Then, for all $i$ (see, e.g.,~\cite{nedic_T-AC-private}):
\begin{equation}
\label{eqn-opt-gap-standard-nondif}
f\left( x_{i,\mathrm{ra}}^{(k)}\right) - f^\star
 \leq
 O\left( \frac{1}{\alpha\,k}\right) + O\left( \alpha\right).
\end{equation}
For method~\eqref{eqn-update-one-node}, we show the following.
Assume that activation probability $p_k = 1-u_k$, $u_k \geq 0$, $\forall k,$
satisfies that:
\begin{equation}
\label{eqn-nondif-cond-u-k}
S_u:= \sum_{k=0}^\infty \sqrt{u_k} < \infty.
\end{equation}
Then, for all~$i$, for all $k=1,2,...$:
{\allowdisplaybreaks{
\begin{eqnarray}
&\,&\mathbb E \left[ f(\overline{x}_{\mathrm{ra}}^{(k)}) - f^\star \right] \leq
\frac{4 N D^2}{2 \alpha k} + \frac{2 \sqrt{2} \,\sqrt{N} \,D\,\sqrt{\mathcal{C}_e}\,S_u}{k}
\nonumber \\
&+&
\hspace{-3mm}{ \alpha} G_{\Psi}^2+
{2 \alpha} \mathcal{C}_e + \frac{\alpha N G^2}{2(1-\lambda_2(C))}
+ \frac{3\alpha N G^2}{p_{\mathrm{min}}(1-\beta)} ,
\label{eqm-nound-nondiff-claim}
\end{eqnarray} }}
where $G_{\Psi}^2$ $:=2$ $ N G^2 $ $+ \frac{18 N G^2 }{(p_{\mathrm{min}})(1-\beta)^2}$.
 Therefore, as long as $p_k$
 converges to one sufficiently fast
 (per condition \eqref{eqn-nondif-cond-u-k} it suffices to
 have, e.g., $p_k = 1 - \frac{1}{(k+1)^{2+\zeta}}$, $\zeta>0$
  arbitrarily small), the
  idling schedule does not violate the
  $O\left( \alpha + \frac{1}{\alpha\,k}\right)$~bound.

\subsection{Quantifying reduction in total cost}
\label{subsection-quantifying-gains}
Although Theorem~3 demonstrates that
the proposed method achieves practically the same
convergence factor (in terms of iterations~$k$)
 as the standard distributed gradient method,
 the Theorem does not
  explicitly quantify the cost reduction needed for achieving a
  prescribed $\epsilon$-accuracy.
  Quantifying this in full generality is very challenging.
 We pursue here the special case of quadratic costs
 with identity Hessians.

\textbf{Setting}. We let $N \geq 2$, and let
 $f_i: \,{\mathbb R}^d \rightarrow \mathbb R$
  be $f_i(x) = \frac{1}{2}\|x-b_i\|^2$,
  $i=1,...,N$, where the $b_i$'s are constant vectors in ${\mathbb R}^d$.
   Note that $\mu=L=1$ and
   $x^\star =\frac{1}{N}\sum_{i=1}^N b_i$.
     Note that this is a rather simple problem,
     where the optimal solution $b^\star$ can be obtained
     if each node solves its own optimization problem, and then the results are averaged,
     e.g., through a consensus algorithm.
     However, it is very useful to illustrate
     in a clear way the cost savings of the proposed method.
   Denote by
   $b^\star:=\mathbf 1 \otimes x^\star$ and $b:=(b_1^\top,...,b_N^\top)^\top
    \in {\mathbb R}^{Nd}$.
    For simplicity, we consider
    equal weights $C_{ij}=c_0$,
    for all $\{i,j\} \in E$,
    so that weight matrix
    $C = I - c_0 \,\mathcal L$, where $\mathcal L$
     is the (un-normalized) zero-one graph Laplacian matrix.
     Denote by $\lambda_i(\mathcal L)$
      the $i$-th \emph{smallest} eigenvalue of
      $\mathcal L$, $i=1,...,N$ (as is common with the Laplacians).
      Then, for $c_0 \leq \lambda_N(\mathcal L)$,
      we have $\|C-J\| = 1-c_0\,\lambda_2(\mathcal L)$.
       From now on, we write simply
       $\lambda_i = \lambda_i(\mathcal L)$.
       Denote by
        $R_{\mathrm{sp}}: = \left\|\left[\,(I-J) \otimes I  \,\right]b\right\|$,
        and by $R_{0}:= \| x^{(0)} - b^\star\|$.
        Quantity $R_{\mathrm{sp}}$
        measures how spread are the $b_i$'s,
        i.e., how the $b_i$'s (minimizers of the individual $f_i$'s)
         are far apart from solution $x^\star = \frac{1}{N}\sum_{i=1}^N b_i$.
          With the proposed method,
           we set
           $p_k = 1-\frac{1}{2}\delta^{k+1}$,
            $k=0,1,...,$
            with~$\delta = 1-\alpha\,\theta$, $\theta \in (0,1/\alpha]$.
            (This is a slightly different choice from
            one considered in the rest of the paper.)
We consider as an error metric the norm of
the mean distance to the solution: $\left\| \mathbb E \left[ x^{(k)}\right] - b^\star\right\|$.
 (The expectation here has no significance
 for the standard distributed gradient method as it is deterministic.)
 This is not a very strong metric, but
 nonetheless it allows to derive neat expressions.
 We denote the latter quantity
 with the standard distributed gradient method
 by $\xi^{(k)}$
  and with the proposed method by~$\chi^{(k)}$.

\textbf{Intermediate results}. We derive the following upper bounds on
$\xi^{(k)}$ and $\chi^{(k)}$, respectively. For all $k=3,4,...$,
there holds:
\begin{eqnarray}
\label{eqn-bounds-xi}
\xi^{(k)} &\leq& \xi_{\mathrm{ub}}^{(k)}:=
(1-\alpha)^k \,R_0 + \alpha\,R_{\mathrm{sp}}(N-1) \\
&\times&
\frac{ 1-(1-\alpha-c_0\,\lambda_2)^k       }{ c_0\lambda_2+\alpha     } \nonumber \\
\label{eqn-bounds-chi}
\chi^{(k)} &\leq& \chi_{\mathrm{ub}}^{(k)}:=
(1-\alpha)^k \,R_0 + \alpha\,R_{\mathrm{sp}}(N-1) \\
&\times&
(\,\frac{ 1}{ c_0\lambda_2(1-\delta^{k/2})+\alpha     }
\nonumber\\
&+&  \frac{ \left(1-\alpha-c_0\,\lambda_2(1-\delta)\right)^{(k-1)/2}}
{ c_0\lambda_2(1-\delta)+\alpha     } \,) .\nonumber
\end{eqnarray}

\textbf{Results}. Based on the above inequalities, we derive the following result.
 Let the desired accuracy be $\epsilon$, i.e., we want that:
 $\xi_{\mathrm{ub}}^{(k)} \leq \epsilon$ and $\chi_{\mathrm{ub}}^{(k)}\leq \epsilon$.
 Then, for $\alpha = \frac{c_0\,\lambda_2\,\epsilon}{2 (N-1)R_{\mathrm{sp}}}$
  and $\theta > \frac{1}{c_0\,\lambda_2}$, after:
   $
  K_{\epsilon} = \frac{R_{\mathrm{sp}} (N-1)}{c_0 \lambda_2 \epsilon} 2\mathrm{ln}\left( \frac{2 R_0}{\epsilon}\right)
   $
 iterations, we have that
 $
 \xi_{\mathrm{ub}}^{(k)} = \epsilon(1+o(\epsilon)),
  $ and $\chi_{\mathrm{ub}}^{(k)} = \epsilon(1+o(\epsilon)),
  $
 i.e., both algorithms achieve the same error $\epsilon$ after the same number of
 iterations $K_{\epsilon}$ (up to lower orders in $\epsilon$). Therefore,
 the proposed method achieves savings in total cost (per node) equal to:
 $
 K_{\epsilon} - \sum_{k=0}^{K_{\epsilon}}p_k ,
 $
 which is approximately
  $
 \frac{1}{2 \alpha \theta} = \frac{(N-1)R_{\mathrm{sp}}}{c_0 \lambda_2 \epsilon}\,\frac{1}{\theta}.
  $
   It is worth noting that, while these worst-case savings
   on the special quadratic costs with identity Hessians
   may not be very large, simulations on
   the more generic costs (with non-unity condition
   numbers $L/\mu$) and real world data demonstrate
    large savings (as presented in Section~V).
 %

\vspace{-3mm}

\section{Conclusion}
\label{section-conclusion}
We explored the effect of two sources of redundancy
with distributed projected gradient algorithms.
The first redundancy, well-known in the literature
on distributed multi-agent optimization, stems from the fact that
not all inter-neighbor links need to be utilized
at all iterations for the algorithm to converge.
 The second redundancy, explored before
only in centralized optimization, arises when
we minimize the sum of cost functions, each summand corresponding to a distinct data sample.
 In this setting, it is known that
performing a gradient method with an
appropriately increasing sample size
can exhibit convergence properties
that essentially match the properties of
a standard gradient method, where the full sample
size is utilized at all times.
 We simultaneously explored the two sources of redundancy
  for the first time to develop a novel
  distributed gradient method.
  With the proposed method, each node, at each iteration~$k$, is active with
  a certain probability $p_k$, and is idle with probability $1-p_k$,
  where the activation schedule is independent
  across nodes and across iterations.
    Assuming that the nodes' local costs
   are strongly convex and have Lipschitz continuous
   gradients, we showed that the proposed
   method essentially matches the
   linear convergence rate (towards
   a solution neighborhood) of the standard
   distributed projected gradient method,
   where all nodes are active at all iterations.
   Simulations on $l_2$-regularized
   logistic losses, both on real world and synthetic data sets,  demonstrate that
    the proposed method significantly reduces
    the total communication and computational cost
    to achieve a desired accuracy, when compared with the
    standard distributed gradient method, and it exhibits
    strong resilience to the effects of asynchrony.

    \vspace{-3mm}

\vspace{-0cm}
\section*{Appendix}

\subsection{Proof of Lemma~5}
\begin{IEEEproof} Consider \eqref{eqn-update-one-node}, and denote by:
{\allowdisplaybreaks{
\begin{eqnarray*}
 y_i^{(k)} :=
    \sum_{j \in \Omega_i\cup \{i\}} W_{ij}^{(k)} x_j^{(k)} - \frac{\alpha z_i^{(k)}}{p_k} \nabla f_i(x_i^{(k)}).
  \end{eqnarray*}}}
   Also, let $\epsilon^{(k)}=(\epsilon_1^{(k)},...,\epsilon_N^{(k)})^\top$,
 where $\epsilon_i^{(k)}:= \mathcal{P}_{\mathcal X}\left\{ y_i^{(k)}\right\} - y_i^{(k)}$.
  Then,~\eqref{eqn-update-compact} can be written in the following equivalent form:
 \begin{equation}
 \label{eqn-update-compact-2}
 x^{(k+1)} = W^{(k)}\,x^{(k)} - \frac{\alpha}{p_k} \left(\,\nabla F(x^{(k)}) \odot z^{(k)}\,\right) + \epsilon^{(k)}.
 \end{equation}
We first upper bound $\|\epsilon^{(k)}\|$.
 Consider $\epsilon_i^{(k)}$.
  We have:
  {\allowdisplaybreaks{
  {\small{
  \begin{eqnarray}
  |\epsilon_i^{(k)}| &=&
  \left| \mathcal{P}_{\mathcal X}\left\{\sum_{j \in \Omega_i\cup \{i\}} W_{ij}^{(k)} x_j^{(k)} - \frac{\alpha z_i^{(k)}}{p_k} \nabla f_i(x_i^{(k)}) \right\} \right. \nonumber \\
  &-& \left.  \left(
  \sum_{j \in \Omega_i\cup \{i\}} W_{ij}^{(k)} x_j^{(k)} - \frac{\alpha z_i^{(k)}}{p_k} \nabla f_i(x_i^{(k)})
  \right)  \right| \nonumber \\
  \label{eqn-chain-1}
  &=&
  \left| \mathcal{P}_{\mathcal X}\left\{\sum_{j \in \Omega_i\cup \{i\}} W_{ij}^{(k)} x_j^{(k)} - \frac{\alpha z_i^{(k)}}{p_k} \nabla f_i(x_i^{(k)}) \right\} \right. \nonumber \\
  &-& \left.  \mathcal{P}_{\mathcal X}\left\{
  \sum_{j \in \Omega_i\cup \{i\}} W_{ij}^{(k)} x_j^{(k)} \right\} + \frac{\alpha z_i^{(k)}}{p_k} \nabla f_i(x_i^{(k)})
   \right| \\
   \label{eqn-chain-2}
   &\leq&
    \left| \mathcal{P}_{\mathcal X}\left\{\sum_{j \in \Omega_i\cup \{i\}} W_{ij}^{(k)} x_j^{(k)} - \frac{\alpha z_i^{(k)}}{p_k} \nabla f_i(x_i^{(k)}) \right\} \right. \nonumber \\
   &-& \left.  \mathcal{P}_{\mathcal X}\left\{
  \sum_{j \in \Omega_i\cup \{i\}} W_{ij}^{(k)} x_j^{(k)} \right\}\right| \nonumber \\
   &+& \left|\frac{\alpha z_i^{(k)}}{p_k} \nabla f_i(x_i^{(k)})
   \right| \\
   \label{eqn-chain-3}
   &\leq&
   \frac{2 \alpha}{p_k} |\nabla f_i(x_i^{(k)})| \leq \frac{2 \alpha \,G}{p_{\mathrm{min}}}.
  \end{eqnarray}}}}}
Equality \eqref{eqn-chain-1}
holds because
$\sum_{j \in \Omega_i\cup \{i\}} W_{ij}^{(k)} x_j^{(k)} \in \mathcal X$,
as a convex combination of the $x_j^{(k)}$
 that belong to $\mathcal X$ by construction, and due to
 convexity of $\mathcal X$.
 Inequality~\eqref{eqn-chain-2}
  is by the triangle inequality.
  Finally,~\eqref{eqn-chain-3}
   is by the non-expansiveness
   property of the Euclidean projection:
   $|\mathcal{P}_{\mathcal X}\{u\} - \mathcal{P}_{\mathcal X}\{v\} |
   \leq |u-v|$, $\forall u,v \in \mathbb R$.
Therefore, we obtain the following bound on $\|\epsilon^{(k)}\|$:
\begin{equation}
\label{eqn-bound-epsilon}
\|\epsilon^{(k)}\| \leq \frac{2 \sqrt{N}\,\alpha \,G}{p_{\mathrm{min}}}.
\end{equation}
We now return to~\eqref{eqn-update-compact-2}. Note that $\widetilde{x}^{(k)}=(I-J)x^{(k)}$. Also,
$W^{(k)} J = J W^{(k)} = J$, by Lemma~\ref{lemma-matrices-W-k}~(b). Thus, we have that:
 $(I-J)W^{(k)} = W^{(k)} - J.$
 Also, $(W^{(k)} - J)(I-J) = W^{(k)} - J - J W^{(k)} + J^2 = W^{(k)} - J $,
 because $J W^{(k)} =J$ and $J^2=J$.
 Using the latter, and multiplying \eqref{eqn-update-compact-2}
 from the left by $(I-J)$, we obtain:
  \begin{eqnarray}
 &\,&\widetilde{x}^{(k+1)} = (W^{(k)}-J)\,\widetilde{x}^{(k)} \nonumber \\
 &-&  \frac{\alpha}{p_k} (I-J)\left(\,\nabla F(x^{(k)}) \odot z^{(k)}\,\right)
 \label{eqn-update-compact-3}
 + (I-J)\epsilon^{(k)}.
 \end{eqnarray}
 Taking the norm and using sub-additive and sub-multiplicative properties:
 {\allowdisplaybreaks{
\begin{eqnarray}
 \|\widetilde{x}^{(k+1)}\| &\leq& \|W^{(k)}-J\|\,\|\widetilde{x}^{(k)}\| \nonumber \\
 &+& \frac{\alpha}{p_k} \|(I-J)\left(\,\nabla F(x^{(k)}) \odot z^{(k)}\,\right)\| \nonumber \\
 &+& \|(I-J)\epsilon^{(k)}\|.  \label{eqn-update-compact-4}
 \end{eqnarray}}}
 It is easy to see that
 $\|\left(\,\nabla F(x^{(k)}) \odot z^{(k)}\right)\| \leq \sqrt{N} G$.
 Hence, using the sub-multiplicative property of norms and
 the fact that $\|I-J\|=1$, there holds:
 $\|(I-J)\left(\,\nabla F(x^{(k)}) \odot z^{(k)}\,\right)\| \leq \sqrt{N} G$.
 Similarly, from~\eqref{eqn-bound-epsilon}:
 $\|(I-J)\epsilon^{(k)}\| \leq \frac{2 \sqrt{N}\,\alpha \,G}{p_{\mathrm{min}}}$.
 Combining the latter conclusions with~\eqref{eqn-update-compact-4}:
\begin{equation}
 \label{eqn-update-compact-5}
 \|\widetilde{x}^{(k+1)}\| \leq \|W^{(k)}-J\|\,\|\widetilde{x}^{(k)}\| + \frac{3 \sqrt{N}\,\alpha \,G}{p_{\mathrm{min}}}.
 \end{equation}
 Squaring the latter inequality, we obtain:
 {\allowdisplaybreaks{
 \begin{eqnarray}
 \|\widetilde{x}^{(k+1)}\|^2 &\leq& \|W^{(k)}-J\|^2\,\|\widetilde{x}^{(k)}\|^2 \nonumber \\
 &+& \frac{6 \sqrt{N}\,\alpha \,G}{p_{\mathrm{min}}}\,
 \|W^{(k)}-J\|\,\|\widetilde{x}^{(k)}\| \nonumber\\
 &+&
  \label{eqn-update-compact-6}
 \left( \frac{3 \sqrt{N}\,\alpha \,G}{p_{\mathrm{min}}}\right)^2.
 \end{eqnarray}}}
 Taking expectation, using the independence of
 $W^{(k)}$ and $\widetilde{x}^{(k)}$,
 and the inequality $\mathbb E [|u|] \leq \left( \mathbb E [u^2]\right)^{1/2}$, for
 a random variable $u$, we obtain:
 {\allowdisplaybreaks{
 {\small{
  \begin{eqnarray}
 &\,& \mathbb E \left[ \|\widetilde{x}^{(k+1)}\|^2 \right]
 \leq \mathbb E \left[\|W^{(k)}-J\|^2\right]\,
 \mathbb E \left[\|\widetilde{x}^{(k)}\|^2\right] \nonumber\\
 &\,& \hspace{4mm}+\frac{6 \sqrt{N}\,\alpha \,G}{p_{\mathrm{min}}}\,
 \left(\mathbb E \left[ \|W^{(k)}-J\|^2\right]\right)^{1/2}\,
 \left(\mathbb E \left[\|\widetilde{x}^{(k)}\|^2\right]\right)^{1/2} \nonumber \\
 &\,&\hspace{4mm}+
  \label{eqn-update-compact-7}
 \left( \frac{3 \sqrt{N}\,\alpha \,G}{p_{\mathrm{min}}}\right)^2.
 \end{eqnarray}
 }}}}
Denote by $\xi^{(k)}:=\left(\mathbb E \left[\|\widetilde{x}^{(k)}\|^2\right]\right)^{1/2}$.
 Applying Lemma~\ref{lemma-matrices-W-k}~(d),
 writing the right-hand side of \eqref{eqn-update-compact-7}
  as a complete square, and taking the square root of the resulting inequality:
\begin{equation}
 \label{eqn-update-compact-10}
 \xi^{(k+1)} \leq \beta\,\xi^{(k)} + \frac{3 \sqrt{N}\,\alpha \,G}{p_{\mathrm{min}}}.
 \end{equation}
 Unwinding recursion~\eqref{eqn-update-compact-10},
 and using the bound $1+\beta+...+\beta^k \leq \frac{1}{1-\beta}$,
 we obtain that, for all $k=0,1,...$, there holds:
 $
\xi^{(k)} $ $\leq $ $\frac{3 \alpha \sqrt{N} G}{p_{\mathrm{min}} (1-\beta)}.
 $
 Squaring the last inequality, the desired result follows.
\end{IEEEproof}

\subsection{Proof of result~(36) in Subsection~{VI-A}}
Assume for simplicity that $d=1$ but the
proof extends to a generic $d>1$.
Let $x^\bullet$ be a solution of~\eqref{eqn-opt-prob-penalty}
(which exists as $\Psi_{\alpha}:\,{\mathbb R}^{N} \rightarrow \mathbb R$ is continuous and
constraint set $\mathcal X$ is compact.)
 The update of the proposed method can be written as:
  \begin{eqnarray}
   \label{eqn-update-penalty-nondif}
   x^{(k+1)} =
     \mathcal{P}_{\mathcal X^N} \left\{x^{(k)} - \alpha \left[\,g_{\Psi}^{(k)} + e^{(k)}\,\right]\right\}. \nonumber
   \end{eqnarray}
Here, $g_{\Psi}^{(k)}$ is a subgradient of $\Psi_{\alpha}$
 at $x^{(k)}$ which equals:
 \begin{eqnarray*}
 g_{\Psi}^{(k)} &=& g_F^{(k)} + \frac{1}{\alpha}(I-C)x^{(k)} \\
 &=& g_F^{(k)} + \frac{1}{\alpha}(I-C)\widetilde{x}^{(k)},
 \end{eqnarray*}
where $g_F^{(k)} = (g_1^{(k)},...,g_N^{(k)})^\top$,
and $g_i^{(k)}$
 is a subgradient of $f_i$ at $x_i^{(k)}$.
 Also, recall $e^{(k)}$ from~(14).
 Note that Lemmas~5 and 9 continue to hold here as well,
 and therefore, it is easy to show that, for all $k=0,1,...$:
\begin{equation}
\label{eqn-subgrad-bnd-nondiff}
\mathbb E \left[ \|g_{\Psi}^{(k)}\|^2\right] \leq
G_{\Psi}^2:=2 N G^2 + \frac{18 N G^2 }{(p_{\mathrm{min}})(1-\beta)^2}.
%
\end{equation}
Now, a standard analysis of
projected subgradient methods, following, e.g.,~\cite{NedicSubgradient}, gives:
\begin{eqnarray*}
\|x^{(k+1)} - x^\bullet\|^2
&\leq&
\|x^{(k)} - x^\bullet\|^2 \\
&-& 2 \alpha \left( x^{(k)} - x^\bullet\right)^\top (g_{\Psi}^{(k)}+e^{(k)})
\\
&+&
\alpha^2 \|g_{\Psi}^{(k)} + e^{(k)}\|^2.
\end{eqnarray*}
Using $\left\| x^{(k)} - x^\bullet \right\| \leq 2 \sqrt{N} D$,
and
\[
\Psi_{\alpha}(x^\bullet)
 \geq
\Psi_{\alpha}(x^{(k)}) + (g_{\Psi}^{(k)})^\top (x^\bullet-x^{(k)}),
\]
we further obtain:
\begin{eqnarray*}
&\,&\|x^{(k+1)} - x^\bullet\|^2
\leq
\|x^{(k)} - x^\bullet\|^2
- 2 \alpha ( \Psi_{\alpha}(x^{(k)}) \\
&\,& -\Psi_{\alpha}(x^\bullet))
+
4\alpha \sqrt{N}D\|e^{(k)}\|
+ 2\alpha^2 \|g_{\Psi}^{(k)}\|^2 +2\alpha^2 \|e^{(k)}\|^2.
\end{eqnarray*}
Summing the above inequality for
$k=0,...,K-1$,
dividing the resulting inequality by $K$,
and using~\eqref{eqn-subgrad-bnd-nondiff}, we obtain:
{\allowdisplaybreaks{
\begin{eqnarray*}
&\,&\frac{2 \alpha}{K} \sum_{k=0}^{K-1} \left(\Psi_{\alpha}(x^{(k)}) - \Psi_{\alpha}(x^\bullet)\right)\\
&\leq&
\frac{\|x^{(0)}-x^\bullet\|^2}{K} + \frac{4 \alpha \sqrt{N} D}{K}\sum_{k=0}^{K-1}\|e^{(k)}\|
\\
&+&
\frac{2 \alpha^2 }{K}\sum_{k=0}^{K-1}\|g_{\Psi}^{(k)}\|^2 +
\frac{2 \alpha}{K}\sum_{k=0}^{K-1}\|e^{(k)}\|^2.
\end{eqnarray*}}}
Consider the running average
$x_{\mathrm{ra}}^{(K)}:=\frac{1}{K}
\sum_{k=0}^{K-1}x^{(k)}$.
Using convexity of~$\Psi_{\alpha}$,
applying~\eqref{eqn-subgrad-bnd-nondiff},
and taking expectation:
\begin{eqnarray}
\label{eqm-nound-nondiff-proof}
&\,&\mathbb E \left[ \Psi_{\alpha}(x^{(K)}) - \Psi_{\alpha}(x^\bullet)\right]\\
&\leq&
\frac{4 N D^2}{2 \alpha K} + \frac{2 \sqrt{N} D}{K}\sum_{k=0}^{K-1}\mathbb E [\|e^{(k)}\|]
\nonumber \\
&+&
{\alpha } G_{\Psi}^2+
\frac{\alpha}{K}\sum_{k=0}^{K-1}\mathbb E [\|e^{(k)}\|^2]. \nonumber
\end{eqnarray}
Next, note that
 $\mathbb E [\|e^{(k)}\|]
 \leq \sqrt{2\mathcal{C}_e} \sqrt{u_k}$,
 and $\mathbb E [\|e^{(k)}\|^2]
  \leq 2 \mathcal{C}_e u_k$,
  where we recall $p_k=1-u_k$.
  Applying
  the latter bounds on~\eqref{eqm-nound-nondiff-proof},
  Using the facts that
  $u_k \geq 1$, for all $k$, and
  that $S_u:=\sum_{k=0}^\infty \sqrt{u_k} $,
  we obtain:
  \begin{eqnarray}
\label{eqm-nound-nondiff-proof-2}
&\,&\mathbb E \left[ \Psi_{\alpha}(x_{\mathrm{ra}}^{(K)}) - \Psi_{\alpha}(x^\bullet)\right]\\
&\leq&
\frac{4 N D^2}{2 \alpha K} + \frac{2 \sqrt{2}  \,\sqrt{N} \,D\,\sqrt{\mathcal{C}_e}\,S_u}{K}
\nonumber \\
&+&
{ \alpha } G_{\Psi}^2+
{2 \alpha} \mathcal{C}_e. \nonumber
\end{eqnarray}
Applying~(10) to $x_{\mathrm{ra}}^{(k)}$.
defining  $\overline{x}_{\mathrm{ra}}^{(k)}:=
\frac{1}{N}\sum_{i=1}^N  x_{i,\mathrm{ra}}^{(k)}$,
 and taking expectations, it follows that:
{\allowdisplaybreaks{
\begin{eqnarray}
\label{eqm-nound-nondiff-proof-2}
&\,&\mathbb E \left[ f(\overline{x}_{\mathrm{ra}}^{(K)}) - f^\star \right]\\
&\leq&
\frac{4 N D^2}{2 \alpha K} + \frac{2 \sqrt{2} \,\sqrt{N} \,D\,\sqrt{\mathcal{C}_e}\,S_u}{K}
\nonumber \\
&+&
{\alpha } G_{\Psi}^2+
{2 \alpha} \mathcal{C}_e + \frac{\alpha N G^2}{2(1-\lambda_2(C))}. \nonumber
\end{eqnarray}}}
Finally, using the same argument as in \cite{arxivVersion}, equation~(22),
we obtain the desired result.

\subsection{Proof of results in Subsection~{VI-B}}
We let $d=1$ for notational simplicity. Consider~$x^{(k)}-b^\star$,
where we recall $b^\star = \frac{1}{N}\sum_{i=1}^N b_i\,\mathbf 1$. Then,
 it is easy to show that, for $k=0,1,...,$ the following recursive equation holds:
 \begin{equation}
 \left( x^{(k+1)} - b^\star\right) = \widetilde{C}
 \left( x^{(k)} - b^\star\right) + \alpha(I-J)b,
 \end{equation}
where $\widetilde{C} = C-\alpha I.$ Therefore,
for $k=1,2,...,$ we have:
\begin{eqnarray}
\label{eqn-proof-special-case-1}
x^{(k)}-b^\star = \widetilde{C}^k(x^{(0)}-b^\star) + \alpha \sum_{t=0}^{k-1}
\widetilde{C}^{k-t} (I-J)b.
\end{eqnarray}
Let ${q}_i$
 denote the $i$-th
 unit-norm eigenvector, and
 ${\lambda}_i$
  the $i$-th eigenvalue
  of Laplacian $\mathcal L$,
  ordered in an ascending order.
  We have that $\lambda_1=0$,
  $\lambda_2>0$, and
  $q_1=\frac{1}{\sqrt{N}}\mathbf 1.$
  Further, note that
$\|\widetilde{C}\| = 1-\alpha$. Then,
there holds:
\[
\widetilde{C}^{k-t} (I-J)b =
\sum_{i=2}^N (1-c_0\lambda_i-\alpha)^{k-t} \widetilde{q}_i \widetilde{q}_i ^\top (I-J)b,
\]
because $q_1^\top (I-J)b=0.$ Therefore, from \eqref{eqn-proof-special-case-1},
we obtain:
\begin{eqnarray*}
\xi^{(k)} \leq (1-\alpha)^k R_0 +
\alpha R_{\mathrm{sp}} (N-1)
\sum_{t=0}^{k-1}
(1-\alpha-c_0 \lambda_2)^{k-t},
\end{eqnarray*}
which yields~(39).

Now, we consider algorithm~\eqref{eqn-update-one-node}.
 Recall quantity~$\chi^{(k)}
  = \|\mathbb E[x^{(k)}] - b^\star\|$.
 Considering the recursive equation
 on $\mathbb E[x^{(k)}] - b^\star$,
 completely analogously to the above, we can obtain:
\begin{eqnarray}
\label{eqn-proof-special-case-chi}
\chi^{(k)} &\leq& (1-\alpha)^k R_0 \\
&+&
\alpha R_{\mathrm{sp}} (N-1)
\sum_{t=0}^{k-1} \prod_{s=t}^{k-1} \left(
1-\alpha-c_0 \lambda_2(1-\delta^{s+1}) \right). \nonumber
\end{eqnarray}
We now upper bound the sum in~\eqref{eqn-proof-special-case-chi}.
 We split the sum into two parts:
 \begin{eqnarray}
\mathcal{S}_1 &=& \sum_{t=0}^{(k-1)/2} \prod_{s=t}^{k-1} \left(
1-\alpha-c_0 \lambda_2(1-\delta^{s+1}) \right)\\
\mathcal{S}_2 &=& \sum_{t=(k-1)/2+1}^{k-1} \prod_{s=t}^{k-1} \left(
1-\alpha-c_0 \lambda_2(1-\delta^{s+1}) \right).
 \end{eqnarray}
  (To avoid notational complications,
we consider odd~$k$ and $k \geq 3$.)
 Next, note that
  \begin{eqnarray*}
\mathcal{S}_1 &\leq& \sum_{t=0}^{(k-1)/2} \left(
1-\alpha-c_0 \lambda_2(1-\delta) \right)^{k-1-t}\\
\mathcal{S}_2 &\leq& \sum_{t=(k-1)/2+1}^{k-1}\hspace{-3mm}  \left(
1-\alpha-c_0 \lambda_2(1-\delta^{k/2}) \right)^{k-1-t}.
 \end{eqnarray*}
From the above bounds, it is easy to show that~\eqref{eqn-bounds-chi} follows.

Now, let $k:=K_{\epsilon} = \frac{\mathrm{ln} \left( \frac{2 R_0}{\epsilon}\right)}{\alpha}$,
and $\alpha = \frac{c_0 \lambda_2 \epsilon}{2 R_{\mathrm{sp}}(N-1)}$.
It is easy to show that,
 for these values,
 quantity $\xi_{\mathrm{ub}}^{(k)}$ in
 (39) is $\epsilon(1+o(\epsilon))$.
  We now show that quantity
  $\chi_{\mathrm{ub}}^{(k)}$ in
 (40) is also~$\epsilon(1+o(\epsilon))$.
 First, note that
 $(1-\alpha)^{K_{\epsilon}}=\frac{\epsilon}{2}(1+o(\epsilon))$.
 Next, consider the term:
 \[
 \frac{\alpha R_{\mathrm{sp}}(N-1)}{c_0\lambda_2(1-\delta^{k/2})+\alpha} \leq
  \frac{\alpha R_{\mathrm{sp}}(N-1)}{c_0\lambda_2(1-\delta^{k/2})}.
 \]
  Note that, for $\theta>0$,
  we have that $\delta^{k/2}
  \sim
   \left(\frac{\epsilon}{2 R_0}\right)^{\theta/2}=o(1)$.
   Therefore, we have that:
\[
  \frac{\alpha R_{\mathrm{sp}}(N-1)}{c_0\lambda_2(1-\delta^{k/2})} = \frac{\epsilon}{2}(1+o(1)).
 \]
   Therefore, we must show that
   the remaining term:
\begin{eqnarray*}
&\,&\frac{ \alpha R_{\mathrm{sp}}(N-1)\left(1-\alpha-c_0\,\lambda_2(1-\delta)\right)^{(k-1)/2}}
{ c_0\lambda_2(1-\delta)+\alpha     } \\
&\leq&
R_{\mathrm{sp}}(N-1)\left(1-\alpha-c_0\,\lambda_2(1-\delta)\right)^{(k-1)/2}
= o(\epsilon).
\end{eqnarray*}
Observe that:
\[
\left(1-\alpha-c_0\,\lambda_2(1-\delta)\right)^{(k-1)/2} \sim \left( \frac{\epsilon}{2 R_0}\right)^{(
1+c_0 \lambda_2 \theta)/2}.
\]
This term is $o(\epsilon)$
  if $\theta>1/(c_0 \lambda_2)$,
 which we assumed, and therefore we
 conclude that $\chi_{\mathrm{ub}}^{(k)}$
  in~(40) is~$\epsilon(1+o(1))$.

\begin{figure}
      \centering
      \includegraphics[height=2.3 in,width=2.9 in]{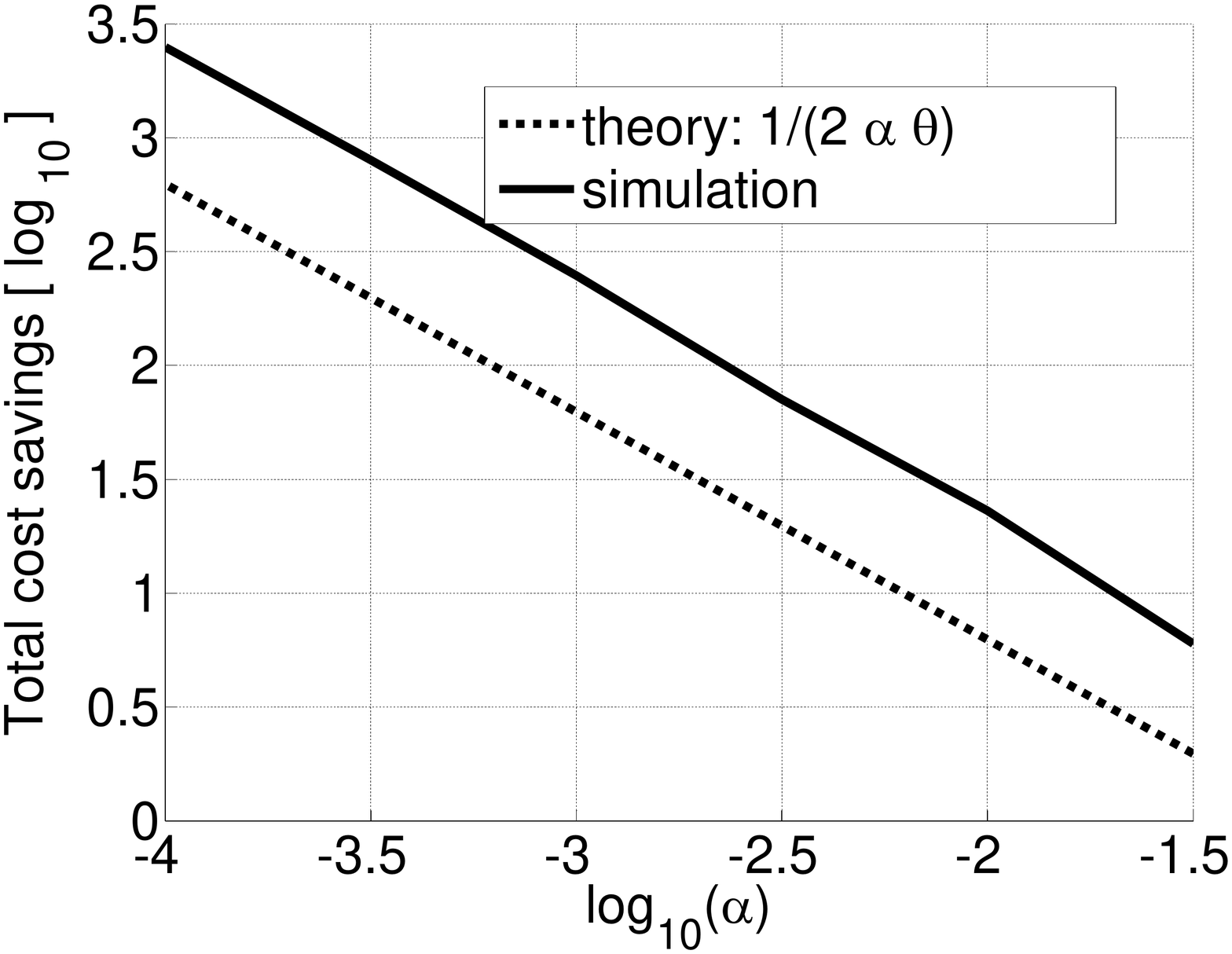}
      \caption{Difference between the total costs of the standard distributed gradient method
      and the proposed method~(in the $\log_{10}$ scale) versus step-size $\alpha$.}
      \label{FigTheoryExample}
\end{figure}

As noted, the established saving of~$\frac{1}{2 \alpha \theta}$ is asymptotic,
  i.e., it holds when the required accuracy~$\epsilon$ (and hence,
  step-size $\alpha$) goes to zero. We now demonstrate by simulation
  (where one sample path is run) that the
  saving of at least~$\frac{1}{2 \alpha \theta}$ holds for finite accuracies also.
   We consider a~$N=4$-node connected network. We set
   $c_0 = 1/(2N)$, and~$\theta=1/(c_0\,\lambda_2(\mathcal{L}))$.
    The quantities $b_i$'s are generated mutually
    independently from the uniform distribution on~$[0,5]$.
   We compare the proposed and standard methods (solution estimates initialized
   with zero at all nodes) in
   terms of the total cost, and the number of iterations,
   needed to reach accuracy~$\epsilon = \epsilon(\alpha) = \frac{2 \alpha (N-1)R_{\mathrm{sp}}}{c_0\,\lambda_2(\mathcal{L})}$,\footnote{Recall that this is the accuracy guaranteed to be achieved
   under step-size~$\alpha$ after~$
  K_{\epsilon} = \frac{R_{\mathrm{sp}} (N-1)}{c_0 \lambda_2 \epsilon} 2\mathrm{ln}\left( \frac{2 R_0}{\epsilon}\right)
   $ iterations.} for the~$\alpha$'s in the range as shown in Table~1. Figure~\ref{FigTheoryExample}
    shows the simulated savings (dotted line) and the savings
    predicted by the (asymptotic) theory (equal to $1/(2\alpha\theta)$).
    We can see that the simulated savings are larger, i.e., they are at least~$1/(2\alpha\theta)$,
    and we can also see that they indeed behave as~$1/\alpha$, as predicted by the theory.

To further corroborate that the gains with the proposed method
       hold in \emph{non-asymptotic} regimes, it is instructive to compare
      it with a naive method which
     stays idle for~$1/(2\alpha\theta)$ iterations,
     and then it continues as the standard distributed gradient method.
      Namely, such a method also has \emph{asymptotic} savings in total cost
      (with respect to the standard method) of~$1/(2\alpha\theta)$
     (as $\alpha \rightarrow 0$), just like the proposed method. However, it is
      clear that such savings are \emph{asymptotic only}, i.e.,
     they do not appear in finite time regimes. That is, with respect to the standard method,
     the naive method only ``shifts'' the error curve along iterations to the right.
     Table~1 shows total costs and iteration costs for $\epsilon$-accuracy
     with the three methods (proposed, standard, and naive). We can see that the proposed method indeed performs
     very differently from the naive method, i.e., it achieves real, non-asymptotic savings.

{\small{
\begin{table*}\centering
\ra{1.3}
\begin{tabular}{@{}rrrrcrrr@{}}\toprule
& $\alpha = 10^{-1.5}$ & $\alpha = 10^{-2}$ & $\alpha = 10^{-2.5}$ & $\alpha = 10^{-3}$ & $\alpha = 10^{-3.5}$ & $\alpha = 10^{-4}$\\ \midrule
\emph{Total cost}\\
$\mathrm{prop.}$ & 106 & 793 & 3965 & 17132 & 68743 & 263479 \\
$\mathrm{stand.}$ & 112& 816& 4036 & 17380& 69540& 265972\\
$\mathrm{naive}$ & 112& 816& 4036 & 17380& 69540& 265972\\
\emph{Number of iterations}\\
$\mathrm{prop.}$ & 28& 203& 1010& 4348& 17377& 66505\\
$\mathrm{stand.}$& 28& 204& 1009& 4345& 17385& 66493&\\
$\mathrm{naive}$ & 29& 210& 1028& 4407& 17582& 67118\\
\bottomrule
\end{tabular}
\caption{Comparison of the proposed method, standard distributed gradient method, and the naive method,
in terms of the required total cost and number of iterations
to reach accuracy~$\epsilon = \frac{2 \alpha (N-1)R_{\mathrm{sp}}}{c_0\,\lambda_2(\mathcal{L})}$.}
\end{table*}}}

\vspace{-0cm}
\bibliographystyle{IEEEtran}
\bibliography{IEEEabrv,bibliography}
\end{document}